\documentclass[reprint,amsmath,amssymb,aps,prb,superscriptaddress]{revtex4-2}
\usepackage{graphicx}
\usepackage{dcolumn}
\usepackage{bm}
\usepackage{color}
\usepackage{blindtext}
\usepackage{hyperref}
\usepackage{amsmath}
\usepackage{enumerate}
\usepackage[final]{changes}
\hypersetup{colorlinks=true, allcolors=blue}

\begin{document}
\title{\deleted[comment={delete}]{Advancing}Quantum Transport Calculations: An Effective Medium Theory with Plane-Wave Basis and PAW Potentials in Eigenstates}

\author{Yi-Cheng Lin}
% \email{b10202018@ntu.edu.tw}
\affiliation{Department of Physics, National Taiwan University, Taipei 106, Taiwan}
\author{Ken-Ming Lin}
% \email{ken.ming.lin@gmail.com}
\affiliation{Department of Electrophysics, National Yang Ming Chiao Tung University, 1001 Daxue Rd., Hsinchu City 300093, Taiwan}
\author{Yu-Chang Chen}
\email{yuchangchen@nycu.edu.tw}
\affiliation{Department of Electrophysics, National Yang Ming Chiao Tung University, 1001 Daxue Rd., Hsinchu City 300093, Taiwan}
\affiliation{Center for Theoretical and Computational Physics, National Yang Ming Chiao Tung University, 1001 Daxue Rd., Hsinchu City 300093, Taiwan}

\begin{abstract}
We present an effective medium theory (EMT-PW) based on density functional theory (DFT) that is implemented in VASP using the projector augmented-wave (PAW) method with a plane-wave basis set. The transmission coefficient, $\tau(E)$, is derived through three complementary approaches: the current density relation $J=nqv$, the field operator method, and the non-equilibrium Green’s function (NEGF) formalism.
%via $\tau = \text{Tr}[\Gamma_L G_r \Gamma_R G_a]$
We compare $\tau(E)$ calculated using EMT-PW with results from NEGF-DFT, based on the NanoDCAL package utilizing a linear combination of atomic orbitals (LCAO) basis set, for both periodic and non-periodic boundary conditions. The minor discrepancies observed are attributed to differences in basis sets, pseudopotentials, and the treatment of lead regions. Notably, the EMT-PW framework avoids the common issue of overcompleteness encountered in non-equilibrium transport theories and allows for the decomposition of the total transmission coefficient into contributions from individual eigenstates,
\replaced{providing an alternative approach to calculate transport properties through nanoscale systems.}{ Furthermore, when combined with an effective gate model, EMT-PW is
shown to be a powerful tool for analyzing current characteristics in nanodevices under
applied gate voltages. By leveraging one-electron wavefunctions in eigenstates, this
method provides a robust foundation for exploring the quantum statistics of electrons
and current quantum correlations within the second quantization framework.}
%, expressed as $\tau(E) = \sum_n \tau_n(E)$, where $\tau_n(E) = \sum_{\mathbf{k}_L} T_{n\mathbf{k}_L}\delta(E - E_{n\mathbf{k}_L})$. The transmission probability density $T_{n\mathbf{k}_L}$ is corresponding to the eigenenergy $E_{n\mathbf{k}}$, associated with the one-electron wavefunction $\psi_{n\mathbf{k}}(\mathbf{r})$.
%providing an alternative approach to calculate transport properties through nanoscale systems.
\end{abstract}
\maketitle

\section{Introduction}\label{sec:intro}
Quantum transport theory was first introduced by Landauer \cite{Landauer1970} and B\"uttiker \cite{Buttiker1986}, with the Landauer formula \cite{Buttiker_book} providing an intuitive understanding of electron transport in two-terminal systems. The non-equilibrium Green’s function (NEGF) method, developed by Keldysh \cite{Keldysh}, offers a more advanced framework for calculating transport currents. Combining NEGF with density functional theory (DFT) enables the study of non-equilibrium atomic systems from first principles. The NanoDCAL simulation package integrates the NEGF method within DFT using a linear combination of atomic orbitals (LCAO) basis \cite{NANODCAL1}. Additionally, incorporating many-body theory with the Lippmann-Schwinger equation into the DFT framework allows for first-principles investigations of shot noise, electron-phonon interactions, and local heating in atomic junctions \cite{doi:10.1021/nl0348544,PhysRevB.67.153304,PhysRevLett.95.166802}. An alternative approach, the wave function matching (WFM) method, is also used to compute electron transport \cite{WFM}. This method was later demonstrated to be equivalent to NEGF \cite{WFM-GF1, WFM-GF2}.

Currently, the most widely used method for studying electronic quantum transport at the atomic scale is the non-equilibrium Green’s function combined with density functional theory (NEGF-DFT) \cite{NANODCAL1,IEEE_NEGF,NEGF_eigenchannels}. 
%{revised in resubmission: 05/12/2025} However, the accuracy of this approach can be affected by the choice of atomic orbitals, and the problem of "overcompleteness" may arise as the number of orbitals increases \cite{Martin2004}. Therefore, exploring alternative methods for calculating the current remains an important and valuable endeavor.
Full self-consistent DFT calculations require specification of the basis. These methods, which utilize a linear combination of localized atomic orbitals (LCAO), are fully ab initio, highly efficient, and often quite accurate. However, this efficiency comes at the cost of generality. Unlike plane-wave basis sets, the atomic orbitals must be carefully selected for each specific system to ensure both accuracy and efficiency. Moreover, attempting to achieve convergence can lead to the issue of "overcompleteness" \cite{Martin2004}. Therefore, exploring alternative methods based on plane-wave bases which serves as a complete basis for calculating current remains an important and valuable pursuit.

This study introduces an effective medium theory derived from wavefunctions obtained using the Vienna Ab initio Simulation Package (VASP) to compute the electric current based on plane-wave bases, denoted as "EMT-PW". The current is advantageously expressed in terms of one-electron wavefunctions, ${\psi }_{n\mathbf{k}}(\mathbf{r})=\frac{1}{\sqrt{\textit{V}}}\sum_{\mathbf{G}}{{\textit{c}}^{ \mathbf{G}}_{n\mathbf{k}}e^{i(\mathbf{k}+\mathbf{G})\cdot \mathbf{r}}}$, which are formulated using a plane-wave basis set. These wavefunctions represent eigenstates within the effective single-particle framework of DFT. The foundation of EMT-PW is the field operator constructed from these wavefunctions: $\hat{\Psi}=\sum_{n{\mathbf{k}}}{{\hat{a}}_{n{\mathbf{k}}}}{\psi }_{n{\mathbf{k}}}(\mathbf{r})$. 

At first glance, the foundation of our EMT-PW theory may seem overly simplistic for accurately describing a true nonequilibrium system. To validate the correctness of the theory, we present a theoretical derivation that links the basis of EMT-PW to the well-established NEGF formalism. This derivation is inspired by the wavefunction matching method \cite{WFM-GF1} and the finite-difference method \cite{PRB-FDM}. We then apply our EMT-PW theory to perform numerical calculations using wavefunctions obtained from VASP for several well-studied systems, including periodic atomic chains \cite{PRB-wannier} and two-dimensional transition metal dichalcogenide (TMD) field-effect transistors (FETs) \cite{FET-AlN}. By comparing our results with those from NEGF-DFT, we demonstrate that EMT-PW provides accurate predictions within the linear response regime. Furthermore, we show that EMT-PW, when combined with the effective gate model within the Landauer formalism, can be employed to analyze atomistic field-effect transistors operating under finite bias. This highlights EMT-PW as a practical and efficient alternative for investigating quantum transport properties in nanoscale systems.  

\begin{figure}
\includegraphics[width=\linewidth]{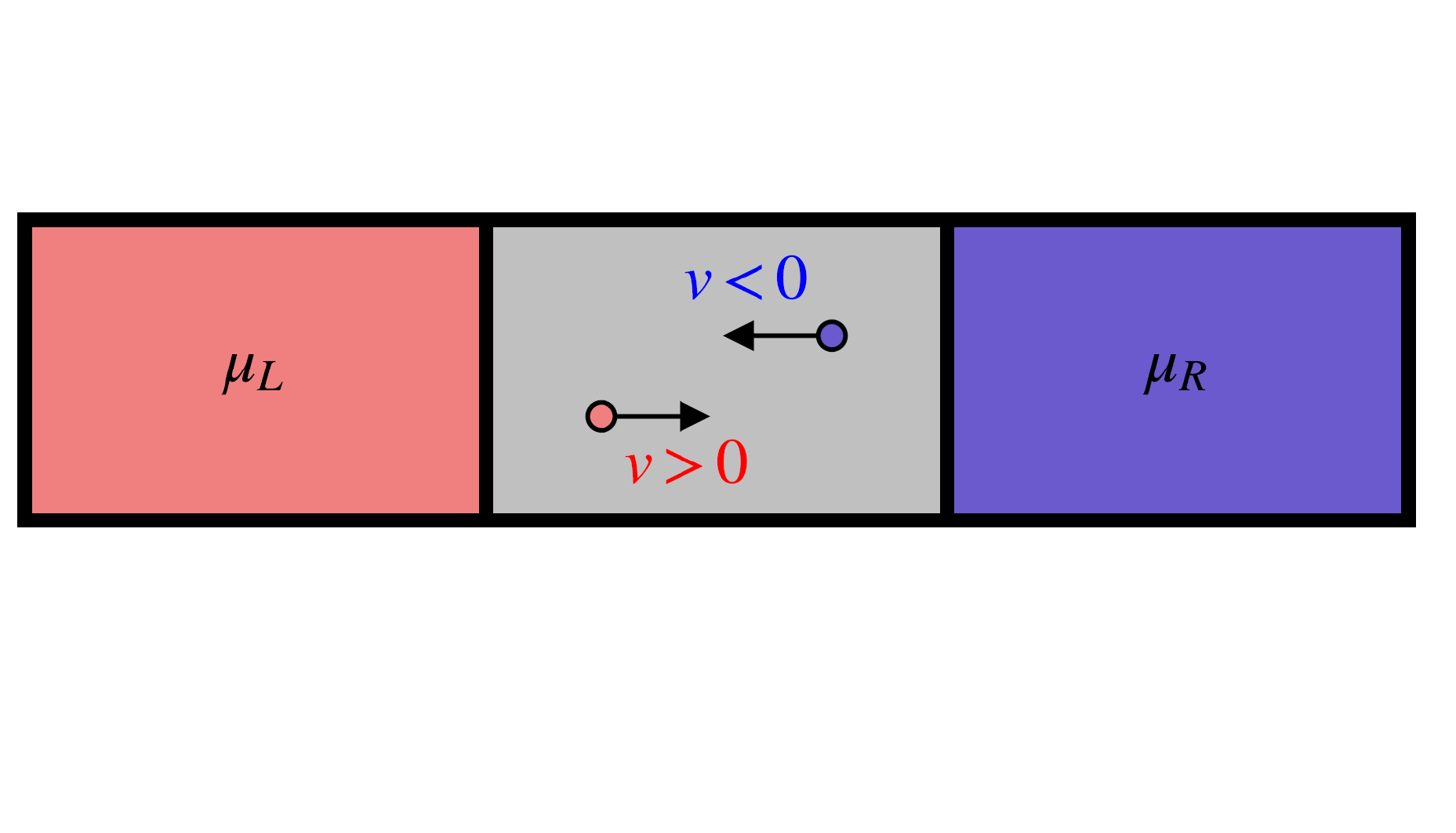}
\caption{\label{fig:system} 
The non-equilibrium system involves different chemical potentials, $\mu_L$ and $\mu_R$, in the left and right reservoirs, respectively.
}
\end{figure}

We begin by introducing the current-density operator methodology. Considering the scheme illustrated in Fig.~\ref{fig:system}, we partition the field operator $\hat{\psi}$ as ${\hat{\psi }} = \hat{\psi}_L + \hat{\psi}_R$, where $\hat{\psi}_L$ and $\hat{\psi}_R$ correspond to electrons incident from the left and right reservoirs, respectively. The field operator $\hat{\psi}_{\alpha}$, with $\alpha=$L or R, are defined as   
\begin{equation}
{\hat{\psi }}_{\alpha}=\sum_{n,{\mathbf{k}}_{\alpha }}\hat{a}_{n \mathbf{k}_{\alpha}} 
\psi_{n \mathbf{k}_{\alpha}}(\mathbf{r}),
\label{eqn:fieldop}
\end{equation}
% Yc Lin: I strongly recommand to omit the spin-index. First, it's consistent with my deviation in section II. Second, in DFT calculation there are three different treatments to the spin degree of freedom:
% 1. spin-degenerated
% 2. spin-polarize
% 3. non-colinear (for spin-orbital interaction, where spin-up/down become ill-defined)
% among them only case 2 possess a well-defined spin-index. Consequently, I believe that omit the spin index will make the deviation more general (as we can absorb the spin-index into "n").
%
% Thus I would like to modify the field operator as follow (I also suggest to drop the time-dependency)
% \begin{align}
% &\hat{\psi} = \hat{\psi}_L + \hat{\psi}_R
% \label{eqn:fieldop},\\
% &\hat{\psi}_{\alpha}=\sum_{n,{\mathbf{k}}_{\alpha }} \hat{a}_{n\mathbf{k}_{\alpha}}\psi_{n\mathbf{k}_{\alpha}}(\mathbf{r}),
% \end{align}
where $\hat{a}_{n \mathbf{k}_{\alpha}}$ is the annihilation operator for an electron in the state characterized by the quantum numbers $n$ and $\Psi_{n\mathbf{k}_\alpha}$ represents the corresponding single-particle wavefunction. We divide the wave vectors $\mathbf{k}$ into two subsets: $\mathbf{k}_L \equiv  \{\mathbf{k}|v_{n\mathbf{k}}^{z}>0 \}$ and $\mathbf{k}_R \equiv  \{\mathbf{k}|v_{n\mathbf{k}}^{z}<0 \}$, corresponding to electrons incident from the left and right reservoirs, respectively. The statistical distribution of these electrons follows the Fermi-Dirac distribution:
\begin{equation}
f_{\alpha}(E_{n\mathbf{k}_{\alpha}})=\frac{1}{e^{{ [E_{n\mathbf{k}_{\alpha}}-{\mu}_{\alpha} ]}/{(k_B T_{\alpha})}}+1}, 
\label{eqn:Fermi-Dirac}
\end{equation}
where $E_{n\mathbf{k}_{\alpha}}$ is the energy of the state, $\mu_\alpha$ is the chemical potential, $T_\alpha$ is the temperature of reservoir $\alpha$ ($\alpha=$L or R). 
%This approach effectively circumvents the issue of "overcompleteness" commonly encountered in non-equilibrium transport theory. <- comment out by Yc Lin on 20250224

We construct the current-density operator using the decomposition $\hat{\psi}={\hat{\psi}}_L+{\hat{\psi}}_R$ in the second-quantized form as \cite{Grosso_ch11,Nazarov_Blanter_2009,PhysRevB.83.035401},
\begin{equation} 
\hat{\mathbf{J}}(\mathbf{r})
\equiv 2 \cdot \frac{e\hbar}{2mi} [{{\hat{\psi }}}^{\dagger }
\nabla {\hat{\psi}}-\nabla 
\hat{\psi}^{\dagger}\hat{\psi} ],
\label{eqn:Jop_def}
\end{equation}
where the factor of 2 accounts for spin degeneracy. The current-density operator can be expressed as
\begin{equation}    
\hat{\mathbf{J}}(\mathbf{r}) = \frac{e\hbar}{mi}   
\sum_{n{n}'} \sum_{\mathbf{k}_{\alpha}\mathbf{k}_{{\alpha}'}}
[\hat{a}^{\dagger}_{n \mathbf{k}_{\alpha}} 
\hat{a}_{{n}'\mathbf{k}_{{\alpha}'}}]
\mathbf{J}_{n{n}',\mathbf{k}_{\alpha} \mathbf{k}_{{\alpha}'}},
\label{eqn:Jop_comp} 
\end{equation}
where
\begin{equation}
\mathbf{J}_{{n}{n}',\mathbf{k}_{{\alpha}} \mathbf{k}_{{\alpha}'} }  
= [\psi_{n\mathbf{k}_{\alpha}}(\mathbf{r})]^{*}\nabla \psi_{{n}'\mathbf{k}_{{\alpha}'}}(\mathbf{r}) -  
\nabla [\psi_{n\mathbf{k}_{\alpha}}(\mathbf{r})]^{*}  \psi_{{n}'\mathbf{k}_{{\alpha}'}}(\mathbf{r}).
\label{eqn:J_comp}  
\end{equation}

The current operator, $\hat{I}$, is obtained by integrating the current-density operator $\hat{\mathbf{J}}(\mathbf{r})$ over the cross-sectional area while averaging along the transport direction $z$:
\begin{equation} 
\hat{I}=\frac{1}{L_z}\int{\hat{\mathbf{J}}(\mathbf{r})d\mathbf{r}}. 
\label{eqn:Iop_def}
\end{equation}
Referring to the scheme in Fig.~\ref{fig:system} and using the quantum statistical expectation value $\langle \hat{a}_{n\mathbf{k_{\alpha}}}^{\dagger} \hat{a}_{n'\mathbf{k_{{\alpha}'}}} \rangle =f_{\alpha}  (E_{n\mathbf{k_{\alpha}}})  \delta_{\alpha {\alpha}'} \delta_{nn'} \delta_{\mathbf{k_{\alpha}} \mathbf{k_{{\alpha}'}}}$, we obtain
\begin{multline}
\langle \hat{I}_z\rangle =\frac{e}{L_z}\sum_{n\mathbf{k}_L}{\{f_R(E_{n \hat{T}[\mathbf{k}_L]})-f_L(E_{n\mathbf{k}_L})\}} 
\\
\cdot\langle \psi_{n\mathbf{k}_{L}}(\mathbf{r}) | \frac{\hat{p}_z}{m} | 
\psi_{n{\mathbf{k}}_{L} }(\mathbf{r}) \rangle,
\label{eqn:Iz}
\end{multline}
where $\hat{p}_{z}=-i\hbar \partial _{z}$ is the z-component of the momentum operator and $\hat{T}[(k_x,k_y,k_z)]=(k_x,k_y,-k_z)$. Note that $\sum_{\mathbf{k}_L}=1$ is due to mode conserving, while $E_{n \hat{T}[\mathbf{k}_L]}=E_{n{\mathbf{k}}_L}$ and $J^{{\sigma}{\sigma}}_{{n}{n}\mathbf{k}_{L}\mathbf{k}_{L}} = -J^{{\sigma}{\sigma}}_{{n}{n}\mathbf{k}_{R}\mathbf{k}_{R}}$ are due to time-reversal symmetry.
Equation (\ref{eqn:Iz}) can be re-expressed in the following form:
\begin{equation}
\langle \hat{I}_z\rangle =\frac{2e}{h}\sum_n{\sum_{{\mathbf{k}}_L}{\{f_R(E_{n{\mathbf{k}}_L}})-f_L( E_{n\hat{T}[{\mathbf{k}}_L]}})\} T_{n\mathbf{k}_L}, 
\label{eqn:Iz_Landauer_K}
\end{equation} 
where
\begin{equation}
 T_{n\mathbf{k}_L}=\frac{h}{L_z}\langle \psi_{n\mathbf{k}_{L}}(\mathbf{r}) | (\hat{p}_{z}/m) | 
\psi_{n{\mathbf{k}}_{L} }(\mathbf{r}) \rangle.
\label{eq:TnK}
\end{equation}
We re-formulate the preceding two equations in a form analogous to the Landauer formula,
\begin{equation}
\langle \hat{I}_z \rangle =\frac{2e}{h}\int (f_R -f_L) \tau (E) dE,
\label{eq:LaudauerE}
\end{equation}
where the total transmission function $\tau (E)$, an analog to that in the non-equilibrium Green’s function (NEGF) formalism, is given by $\tau (E)=\sum_{n} \tau_{n}(E)$ with the mode-resolved transmission function $\tau_n (E)$ defined as 
\begin{equation}
\tau_n (E)= \sum_{\mathbf{k}_L} T_{n\mathbf{k}_L}\delta[E-E_{n\mathbf{k}_L}],    
\label{eqn:tau(E)1}
\end{equation} 
where $T_{n\mathbf{k}_L}$ is the transmission probability density at energy $E_{n\mathbf{k}_L}$, weighted by the Dirac delta function $\delta[E-E_{n\mathbf{k}_L}]$. The information about the density of states (DOS) is inherently included in $\tau_n (E)$, as the partial density of states for electrons in the $n-$th band incident from the right reservoir is given by $D_n (E)= \sum_{\mathbf{k}_L} \delta[E-E_{n\mathbf{k}_L}]$. Thus, $\tau_n (E)$ can be interpreted as the product of the transmission probability density $T_{n\mathbf{k}_L}$ and the corresponding partial density of states, effectively capturing both the likelihood of electron transmission and the availability of states at energy $E_{n\mathbf{k}_L}$.

From an alternative perspective, the current density in a solid-state system can be expressed as
\begin{equation}\label{eqn:J}
\mathbf{J} = \frac{2(-e)}{V}\sum_{n \mathbf{k}} (f_{R}-f_L) \mathbf{v}_{n \mathbf{k}}
\end{equation}
where $\mathbf{v}_{n \mathbf{k}}$ is the velocity of an electron with wavevector $\mathbf{k}$ in the $n-$th band; $V$ is the volume of the system, and the factor of two accounts for spin degeneracy. It is evident that Eq.~(\ref{eqn:J}) is a band-resolved version of the classical expression $J = nqv$, where $q=-e$, $n=N/V$, and the total number of electrons $N$ is given by $N \mapsto 2\sum_{n}\sum_{\bf{k}}$. In the system illustrated in Fig.~\ref{fig:system}, with the transport direction oriented along the $z-$axis, right-going electrons ($v^z_{n \mathbf{k}} > 0$) originate from the left electrode, and their occupation is governed by the Fermi-Dirac distribution $f_L$. Conversely, left-moving electrons ($v^z_{n \mathbf{k}} < 0$) are populated according to the Fermi-Dirac distribution $f_R$ from the right electrode. This viewpoint emphasizes the role of the distribution functions in determining the net current flow, where the imbalance in occupation between left- and right-moving states drives the transport.

%In the framework of Boltzmann transport, we can solve $f_{n \mathbf{k}}$ with Boltzmann equation to obtain the diffusive current. However, in ballistic limit there's no scattering to damp the electron transport, so the procedure described above become invalid.To overcome such issue, we can heuristically assert a physical argument as follow: That is \[f_{n \mathbf{k}} = \begin{cases} f_L(E_{n \mathbf{k}})&\text{if}\quad v^z_{n \mathbf{k}} > 0\\  f_R(E_{n \mathbf{k}})&\text{if}\quad v^z_{n \mathbf{k}} < 0 \end{cases}\] where $f_{L,R}(E) = 1/(e^{(E-\mu_{L, R})/k_BT} + 1)$ are related to the chemical potential $\mu$ of right lead and left lead respectively. 

According to time-reversal symmetry \cite{time_reversal}, each electron state with velocity $v^z_{n \mathbf{k}}$ has a corresponding state with wavevector $\hat{T}[\mathbf{k}]$ and an opposite velocity $v^z_{n \hat{T}[\mathbf{k}]}$. This symmetry allows us to reformulate the summation over states with $v^z_{n \mathbf{k}} > 0$ (or $v^z_{n \mathbf{k}} < 0$) by replacing $v^z_{n \mathbf{k}}$ with $|v^z_{n \mathbf{k}}|/2$ and extending the summation to include all states. Applying this approach to Eq.~(\ref{eqn:J}), we obtain the following expression for the current:
\begin{equation}
\label{eqn:Laudauer2} 
I = \frac{2e}{h} \int dE , (f_L - f_R) \left(\frac{h}{L_z} \sum_{n, \mathbf{k}} \delta(E - E_{n \mathbf{k}}) \frac{|v^z_{n \mathbf{k}}|}{2}\right). 
\end{equation}
To identify the analogous transmission coefficient, we equate Eq.~(\ref{eqn:Laudauer2}) with the Landauer formula,
\begin{equation} I = \frac{2e}{h} \int dE  (f_L - f_R)  \tau(E) \end{equation}
which yields the following expression for the transmission coefficient:
\begin{equation}
\label{eqn:semi-tau} 
\tau(E) = \frac{h}{L_z} \sum_{n, \mathbf{k}} \delta\left[E - E_{n\mathbf{k}}\right] \frac{|v^z_{n \mathbf{k}}|}{2}. 
\end{equation}

The transmission coefficient $\tau(E)$ derived here agrees with the expression in Eq.~(\ref{eqn:tau(E)1}), where we have established the correspondence between $\langle \psi_{n\mathbf{k}}(\mathbf{r}) | \frac{\hat{p}{z}}{m} | \psi_{n\mathbf{k}}(\mathbf{r}) \rangle$ and $v^z_{n \mathbf{k}}$. This correspondence establishes a link between the expectation value of the momentum operator in the $z-$direction and the semiclassical electron velocity.

% Yc Lin: "v^z = p_z / m" is is kind of common sense in my point of view.

This paper aims to compute the transmission coefficient, $\tau(E)$ [Eq.~(\ref{eqn:semi-tau})], analogous to that obtained from NEGF-DFT, using conventional DFT methods, specifically the VASP simulation package in plane-wave basis. We refer to this approach as the "EMT-PW" theory. Notably, EMT-PW circumvents the "overcompleteness" issue commonly encountered in non-equilibrium transport theory. A key advantage of EMT-PW is its ability to distinguish the transmission coefficient in eigenchannels: $\tau(E)=\sum_{n} \tau_n (E)$, where $\tau_n (E)=\sum_{n\mathbf{k}_L}T_{n\mathbf{k}_L}\delta(E-E_{n\mathbf{k}_L})$. The transmission probability density $T_{n\mathbf{k}}$ is defined on the $n-$th energy band $E_{n\mathbf{k}}$ which is associated with the one-electron wavefunction $\psi_{n\mathbf{k}}(\mathbf{r})$. Unlike the transmission coefficient $\tau(E)$ from NEGF-DFT in an open system, which is challenging to resolve contributions from individual electrons in quantum channels, EMT-PW can distinguish contributions from each electron in the eigenstate. As a result, EMT-PW provides a robust framework for investigating the quantum statistics of electrons and the quantum correlations of currents. Furthermore, integrating EMT-PW with the effective gate model \cite{FET-AlN} offers a powerful approach for exploring quantum transport properties in nanojunctions under the influence of a gate field.

% Note that VASP employs the PAW method for the one-electron wavefunctions $\psi_{n\mathbf{k}}$, necessitating an additional linear transformation via the pseudo orbitals \cite{uspp_to_paw}.
In the EMT-PW theory, there are no leads, and electrons are not physically scattered by the central scattering region. The absence of leads makes it challenging to interpret the analogous transmission coefficient, which is derived through direct comparison with the Landauer formula. Therefore, the theoretical section of this paper focuses on the justification of $\tau(E)$ within the EMT-PW framework, specifically Eq.~(\ref{eqn:semi-tau}), or equivalently, Eq.~(\ref{eqn:tau(E)1}). To this end, we re-derive Eq.~(\ref{eqn:semi-tau}) using the NEGF formalism and demonstrate the equivalence between EMT-PW and NEGF-DFT. In addition, we apply the EMT-PW theory to a one-dimensional metallic atomic chain, we successfully reproduce the quantized conductance phenomenon \cite{NEGF-QPC}, similar to results from NEGF-DFT. This comparison validates the accuracy and reliability of our EMT-PW approach.

In Sec.~\ref{sec:theory}, we re-derive  Eq.~(\ref{eqn:semi-tau}), or equivalently Eq.~(\ref{eqn:tau(E)1}) from the NEGF formalism. We demonstrate that, under certain assumptions, the analogous transmission coefficient in the EMT-PW theory is identical to the transmission coefficient obtained from NEGF. In Sec.~\ref{sec:result}, we present numerical results for the analogous transmission coefficients in the EMT-PW framework and compare them with the transmission coefficients obtained from NEGF-DFT. Our findings demonstrate a strong agreement between the two approaches. In Appendix, We provide a detailed explanation of how to calculate Eq.~(\ref{eqn:semi-tau}) or equivalently Eq.~(\ref{eqn:tau(E)1}). The calculations are based on one-electron wavefunctions obtained from VASP, utilizing a plane-wave basis set and the projector-augmented-wave (PAW) method. The VASP parameters and the Gaussian smearing parameters are also presented here.

\section{Theory}\label{sec:theory}
\subsection{Periodic System as Open System}

The non-equilibrium Green’s function (NEGF) formalism is widely used to describe open systems comprising two semi-infinite leads connected to a central scattering region, as depicted in Fig.~\ref{fig:system}. Within the NEGF framework, the current flowing through such a non-equilibrium system is given by \cite{PRB-extended,NANODCAL1}:
\begin{equation*}
    I = \frac{2e}{h}\int dE(f_L - f_R)\tau(E),
\end{equation*}
where the transmission coefficient $\tau(E)$ is expressed as
\begin{equation}
\label{eqn:negf} 
\tau(E) = \text{Tr}\left[\Gamma_L G_r \Gamma_R G_a\right]. 
\end{equation}
Here, $G_r $ and $G_a$ are the retarded and advanced Green’s functions of the central region, respectively, and $\Gamma_{L/R}$ are the broadening matrices for the left and right leads. The broadening matrix is defined as the anti-Hermitian part of the self-energy $\Gamma = i(\Sigma - \Sigma^{\dagger})$, where $\Gamma$  is the self-energy representing the effect of the semi-infinite leads, typically computed from the leads' surface Green’s functions.

However, an alternative viewpoint arises from the established equivalence between the Green’s function method and the wavefunction matching (WFM) method \cite{WFM-GF1,WFM-GF2}. This equivalence suggests that the essential information required to calculate the transmission coefficient lies in the local behavior of the wavefunction at the interface between the leads and the central scattering region. Consequently, direct reliance on the leads’ Green’s functions becomes unnecessary. Instead, a properly constructed wavefunction that accurately mimics the behavior of the infinite lead at the boundary of the scattering region can sufficiently capture the relevant physics.

This insight opens the door to modeling open systems using periodic supercells, where the scattering region and its boundaries are embedded in a periodic framework. By employing well-constructed wavefunctions at the supercell boundaries, it becomes possible to bypass the explicit calculation of self-energies, simplifying the computational approach while retaining accuracy in evaluating transport properties.

As discussed in Ref.~\cite{PRB-wannier}, the transmission coefficient in periodic systems can be computed by constructing Green’s functions using matrix elements defined in the Wannier function basis. This approach enables the calculation of electron transmission through infinitely long periodic chemical chains with high accuracy and efficiency.Moreover, this method is equally applicable to non-periodic junction systems comprising a central scattering region connected to two leads. Such systems can be effectively modeled within a periodic supercell framework, allowing for the accurate evaluation of transport properties without explicitly treating the semi-infinite leads. This approach is demonstrated in Fig.\ref{fig:result}(c)(d) and further discussed in Sec.\ref{sec:result}.

Based on S. Datta's formulation \cite{Datta_2005}, the quantum state $|\Psi\rangle$ of an open system, consisting of a central scattering region connected to two semi-infinite leads, is governed by the following equation:
\begin{equation}\label{eqn:datta}
(E - H - \Sigma_L - \Sigma_R)|\Psi\rangle = |S\rangle,
\end{equation}
where $|S\rangle$ represents the source term (i.e., the incoming wave) that is independent of the central state's evolution, and $\Sigma_{L/R}|\Psi\rangle$ captures the effect of outgoing waves, characterizing how the central state $|\Psi\rangle$ couples to the external leads through the self-energies $\Sigma_L$ and $\Sigma_R$.

To numerically solve Eq.~(\ref{eqn:datta}), the Finite Difference Method (FDM)—commonly used in NEGF implementations \cite{PRB-FDM}—can be employed to project the coupling between the leads and the central region onto real space. In this approach, the total Hamiltonian is discretized along the transport direction (the $z$-axis), resulting in a tridiagonal matrix structure where the only off-diagonal elements arise from the discretization of the kinetic energy term.

This discretized form simplifies the modeling of quantum transport by reducing the lead-central coupling to real-space hopping terms, thereby enabling efficient computation of Green’s functions and transport properties within the NEGF framework:
\begin{align*}
    -\frac{\hbar^2}{2m}\frac{d^2}{dz^2}\Psi(z) &= \frac{\hbar^2}{2m}\frac{2\Psi(z) - \Psi(z + \Delta) - \Psi(z - \Delta)}{\Delta^2}\\
    &= (\frac{\hbar^2}{2m}\frac{2\langle z| - \langle z + \Delta| - \langle z - \Delta |}{\Delta^2})|\Psi\rangle.
\end{align*}
Here $|z\rangle$ is the position eigenstate in matrix mechanics language. Thus, the Hamiltonian can be written as
\[H = \sum_{zz'} |z\rangle H_{zz'} \langle z' | ,\]
where the only non-zero off-diagonal terms are $H_{z, z\pm\Delta} = -\hbar^2/2m\Delta^2$. This allows us to reduce the Schr\"odinger equation $(E - H)|\Psi_{\text{tot}}\rangle = 0$ to an equation for the central region,
\begin{widetext}
\[ [E - H_C + \frac{\hbar^2}{2m\Delta^2}\bigg(|-\Delta \rangle\langle 0| + |0 \rangle\langle -\Delta| + |L_z + \Delta \rangle\langle L_z| + |L_z \rangle\langle L_z + \Delta|\bigg) ]|\Psi\rangle = 0 ,\]
where $L_z$ is the length of the central region in the z-direction, and $H_C = \sum_{zz'\in [0, L_z]} |z\rangle H_{zz'} \langle z' |$ is the central Hamiltonian. To map such an equation to Datta's formulation in Eq.~(\ref{eqn:datta}), we need to ensure that the state on the left side is only projected to the central region. Those projected to the lead region must be moved to the right-hand side and identified as the "source" of the center system.  Hence, we arrive at
\[\bigg[E - H_C - \underbrace{\frac{-\hbar^2}{2m\Delta^2}|-\Delta \rangle\langle 0|}_{\Sigma_L} - \underbrace{\frac{-\hbar^2}{2m\Delta^2}|L_z + \Delta \rangle\langle L_z|}_{\Sigma_R}\bigg]|\Psi\rangle = \underbrace{\frac{-\hbar^2}{2m\Delta^2}\bigg(|0 \rangle\langle -\Delta|\Psi\rangle + |L_z \rangle\langle L_z + \Delta|\Psi\rangle\bigg)}_{|S\rangle} .\]
\end{widetext}
It is worth noting that the right-hand side is only dependent on the wavefunction outside of the center region. Such separation of terms enables us to identify the self-energies,
\begin{equation}\label{eqn:self-energy}
\begin{split}
   &\Sigma_L = \frac{-\hbar^2}{2m\Delta^2} |-\Delta \rangle\langle 0|, \\
   &\Sigma_R = \frac{-\hbar^2}{2m\Delta^2} |L_z + \Delta \rangle\langle L_z| ,
\end{split}
\end{equation}
which will serve as the foundational assumptions for the subsequent mathematical procedures.

\subsection{Transmission Coefficient}
At first appearance, the operator form of self-energy in Eq.~(\ref{eqn:self-energy}) seems hardly illuminated; however, the broadening matrices $\Gamma = i(\Sigma - \Sigma^{\dagger})$ may be reduced to a quite physical form if we use the translation operator \cite{Sakurai} as follows: for the left lead, we rewrite  $|-\Delta\rangle$ by the translation operator $|-\Delta\rangle = e^{-i\hat{p}_z(-\Delta)/\hbar}| 0 \rangle \approx (1+i\hat{p}_z\Delta/\hbar)| 0 \rangle$, and plug in using Eq.~(\ref{eqn:self-energy}). We get
\[\Gamma_L = \frac{\hbar}{2m\Delta} (\hat{p}_z|0\rangle\langle 0| + |0\rangle\langle 0|\hat{p}_z) .\]
Similarly, $|L_z+\Delta\rangle = e^{-i\hat{p}_z\Delta/\hbar}| L_z \rangle \approx (1-i\hat{p}_z\Delta/\hbar)| L_z \rangle$, and we get 
\[\Gamma_R = -\frac{\hbar}{2m\Delta} (\hat{p}_z|L_z\rangle\langle L_z| + |L_z\rangle\langle L_z|\hat{p}_z) .\]
Define the first-quantized version of current density operator as
\begin{equation}\label{eqn:current_operator}
\hat{\mathbf{j}}(\mathbf{r}) = \frac{1}{2m}\big(\ \hat{\mathbf{p}}|\mathbf{r}\rangle\langle\mathbf{r}| + |\mathbf{r}\rangle\langle\mathbf{r}|\hat{\mathbf{p}}\ \big) .
\end{equation}
For details, see Appendix \ref{app:current_density}. In addition, we define the averaged current operator $\hat{\overline{\mathbf{j}}}(z) \equiv \int \text{d}x\text{d}y\ \hat{\mathbf{j}}(\mathbf{r})$. We then can rewrite the broadening matrices in terms of the averaged current operator, which describe the probability "flux" flowing through the left/right cell boundary as
\begin{equation}\label{eqn:gamma}
\Gamma_L = \frac{\hbar}{\Delta}\hat{\overline{j}}_z(0),~~
\text{and} ~~\Gamma_R = -\frac{\hbar}{\Delta}\hat{\overline{j}}_z(L_z).
\end{equation}
Considering the Green's function in the form of
\begin{equation}\label{eqn:G_ra}
G_{r/a} = \sum_{n \mathbf{k}} \frac{|\psi_{n \mathbf{k}}\rangle\langle \psi_{n \mathbf{k}}|}{E \pm i\epsilon - E_{n \mathbf{k}}},
\end{equation}
where $\epsilon \rightarrow 0^+$ , we plug Eq.~(\ref{eqn:gamma}) and Eq.~(\ref{eqn:G_ra}) into the transmission coefficient in NEGF formalism,
\begin{align*}
\tau(E) &= \text{Tr}[\Gamma_L G_r \Gamma_R G_a]\\
&= -\frac{\hbar^2}{\Delta^2}\sum_{nm\mathbf{k}\mathbf{q}}\frac{\langle \psi_{m \mathbf{q}}|\hat{\overline{j}}_z(0)|\psi_{n\mathbf{k}}\rangle\langle \psi_{n\mathbf{k}}|\hat{\overline{j}}_z(L_z)|\psi_{m \mathbf{q}}\rangle}{(E + i\epsilon - E_{n\mathbf{k}})(E - i\epsilon - E_{m \mathbf{q}})}.
\end{align*}
where there is an issue with the minus sign in the above expression, as the transmission coefficient should be positive rather than negative. The negative sign arises from the way self-energy is described, where both terms represent the outflow of probability current, leading to two fluxes flowing in opposite directions. Since we are only concerned with the magnitude of the transmission coefficient, we will discard the unphysical minus sign. 

\begin{figure}
\includegraphics[width=0.9\linewidth]{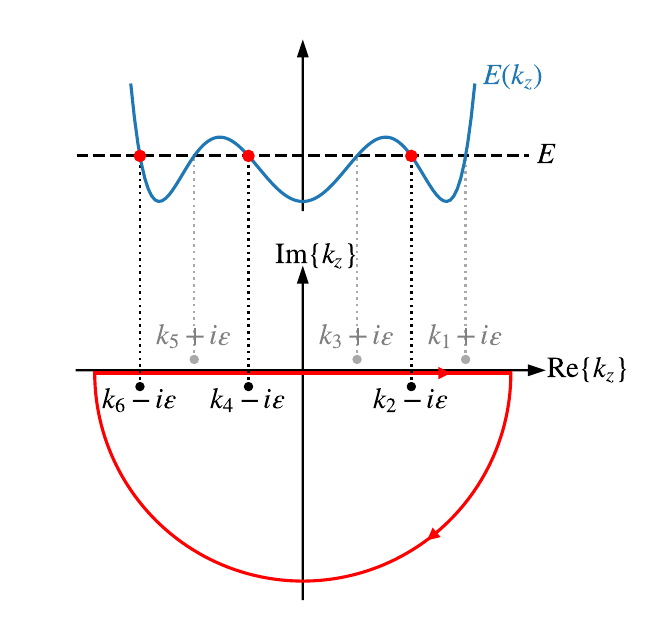}
\caption{\label{fig:complex} 
The band structure and the contour used to evaluate $F_{\mathbf{k}_{\parallel}}(E)$ are shown. It is observed that all $k-$points enclosed inside the contour integral path exhibit negative group velocities.
}
\end{figure}

The Bloch theorem [$\langle \psi_{n\mathbf{k}}|\hat{\overline{j}}_z(L_z)|\psi_{m \mathbf{q}}\rangle = e^{i(q_z - k_z)L_z} \langle \psi_{n\mathbf{k}}|\hat{\overline{j}}_z(0)|\psi_{m \mathbf{q}}\rangle$] and the fact that the current density operator is hermitian allow us to rewrite the transmission coefficient as,
\begin{align}
\tau(E) &= \frac{\hbar^2}{\Delta^2}\sum_{nm\mathbf{k}\mathbf{q}}\frac{ |\langle \psi_{m \mathbf{q}}|\hat{\overline{j}}_z(0)|\psi_{n\mathbf{k}}\rangle |^2 e^{i(q_z - k_z)L_z}}{(E + i\epsilon - E_{n\mathbf{k}})(E - i\epsilon - E_{m \mathbf{q}})}\nonumber\\
&= \frac{\hbar^2}{\Delta^2}\sum_{\mathbf{k}_{\parallel}\mathbf{q}_{\parallel}} |\langle \psi_{m \mathbf{q}}|\hat{\overline{j}}_z(0)|\psi_{n\mathbf{k}}\rangle |^2 F_{\mathbf{k}_{\parallel}}(E) F^*_{\mathbf{q}_{\parallel}}(E)\label{eqn:mid_step},
\end{align}
where we define $F_{\mathbf{k}_{\parallel}}(E)$ as
\[F_{\mathbf{k}_{\parallel}}(E) \equiv \sum_{n, k_z}\frac{e^{-ik_z L_z}}{E + i\epsilon - E_{n\mathbf{k}}}.\]
By relabeling the original reduced zone scheme $(n, k_z)$, where $k_z\in [-\pi/2L_z, \pi/2L_z])$, into an extended zone scheme $k_z\in [-\infty, \infty]$, where the label $n$ is absorbed into $k_z$, the factor $F_{\mathbf{k}_{\parallel}}(E)$ becomes
\begin{equation}\label{eqn:F(E)}
F_{\mathbf{k}_{\parallel}}(E) = \frac{L_z}{2\pi}\int_{-\infty}^{\infty} dk_z\frac{e^{-ik_z L_z}}{E + i\epsilon - E_{\mathbf{k}_{\parallel}}(k_z)}.
\end{equation}
Strictly speaking, $E_{\mathbf{k}_{\parallel}}(k_z)$ is not a continuous function since there may be discontinuities at the zone boundary when a small gap exists. However, because we only care about the value of $F_{\mathbf{k}_{\parallel}}(E)$ around the pole of $E - E_{\mathbf{k}_{\parallel}}(k_z)$, as long as the energy $E$ is far away from the small gaps, it's still proper to regard it as a continuous function. With this assumption, we may expand $E_{\mathbf{k}_{\parallel}}(k_z)$ as a polynomial of $k_z$, and the denominator can be rewritten as,
\begin{widetext}
\[\text{LHS} = E + i\epsilon - E_{\mathbf{k}_{\parallel}}(k_z)
= -\alpha (k_z - k_1 - i\epsilon)(k_z - k_2 + i\epsilon)(k_z - k_3 - i\epsilon)\cdots = \text{RHS}. \]
where we assumed that the energy $E_{\mathbf{k}_{\parallel}}(k_z)$ is positive for $k_z \rightarrow\pm\infty$, so the leading coefficient $\alpha>0$, and $k_i$ are the zeros of the function $(E - E_{\mathbf{k}_{\parallel}})$ ordering as $k_1>k_2>k_3>\cdots$.  Note that for RHS, the sign of $i\epsilon$ are arranged alternatively so that the sign of RHS is consistent with LHS at all poles. For example:

\begin{align*}
&k_z = k_1: &&\text{RHS} = \alpha  (0 + i\epsilon)\underbrace{(k_1 - k_2 + i\epsilon)}_{>0} \underbrace{(k_1 - k_3 - i\epsilon)}_{>0}\cdots  \quad\Rightarrow\quad \text{Im(RHS)}>0 ,\\
&k_z = k_2: &&\text{RHS} = \alpha \underbrace{(k_2 - k_1 - i\epsilon)}_{<0} (0 -i\epsilon)\underbrace{(k_2 - k_3 - i\epsilon)}_{>0}\cdots  \quad\Rightarrow\quad \text{Im(RHS)}>0 ,
\end{align*}
which is consistent with Im(LHS) $ > 0$ as expected. Plug this expression back to the integral in Eq.~(\ref{eqn:F(E)}), and evaluate it by the contour integral shown in lower panel of Fig.~\ref{fig:complex}.
\begin{align*}
F_{\mathbf{k}_{\parallel}}(E) &= \frac{L_z}{2\pi}\sum_{i\in \text{even}} -2\pi i\mathop{\mathrm{Res}}_{k_z = k_i}\frac{e^{-ik_z L_z}}{-\alpha (k_z - k_1 - i\epsilon)(k_z - k_2 + i\epsilon)\cdots}= -iL_z\sum_{i\in \text{even}} \frac{e^{-ik_iL_z}}{\frac{\text{d}\ }{\text{d}k_z}\big(E - E_{\mathbf{k}_{\parallel}}(k_i)\big)} =  \frac{iL_z}{\hbar}\sum_{i\in \text{even}} \frac{e^{-ik_iL_z}}{v^z_{\mathbf{k}_{\parallel}}(k_i)}.
\end{align*}
\end{widetext}
The energy band $E(k_z)$ plotted in the upper panel of Fig.~\ref{fig:complex} clearly shows that all poles $k_i$ encompassed inside the contour (i.e., even $i$) have negative group velocity. Consequently, we can replace the summation $\sum_{i\in \text{even}}$ by $\sum_i \Theta(-v^z(k_i))$, where $\Theta(v)$ is heaviside step function. Now we can see the brilliance of the contour integral as follows: In section \ref{sec:intro}, we used the selective fermi-Dirac distribution for $v^z > 0$ as $f_L$ (and $v^z < 0$ as $f_R$). But here, such selective behavior emerges purely mathematically from the step function $\Theta(-v^z)$ because the contour integral only encloses those states with negative velocity.
%At this point, one may be concerned about why "negative" velocity is special in our deviation, which seems to break the inversion symmetry. In fact, if we rewrite the transmission coefficient as $\tau(E) = \text{Tr}[\Gamma_L G_r \Gamma_R G_a] = \text{Tr}[\Gamma_R G_r \Gamma_L G_a]$ (by taking complex conjugate), then we will get the positive velocity instead. 
No matter which sign of the velocity is chosen, the value of transmission coefficient must remain the same due to the time-reversal symmetry $E(\mathbf{k}) = E(-\mathbf{k})$ \cite{time_reversal}.

Since $k_i$ is the root of $E - E_{\mathbf{k}_{\parallel}}(k_z) = 0$, we can rewrite the formula as
\begin{align*}
F_{\mathbf{k}_{\parallel}}(E) &= \frac{iL_z}{\hbar}\sum_i \frac{e^{-ik_iL_z}}{v^z_{\mathbf{k}_{\parallel}}(k_i)}\Theta(-v^z_{\mathbf{k}_{\parallel}}(k_i))\\
&= \frac{iL_z}{\hbar}\int dE_{\mathbf{k}_{\parallel}} \cdot \frac{e^{-ik_zL_z}}{v^z_{\mathbf{k}_{\parallel}}}\Theta(-v^z_{\mathbf{k}_{\parallel}}) \delta(E - E_{\mathbf{k}_{\parallel}})\\
&=iL_z\int dk_z\cdot e^{-ik_zL_z}\Theta(-v^z_{\mathbf{k}_{\parallel}}(k_z)) \delta(E - E_{\mathbf{k}_{\parallel}}(k_z)),
\end{align*}
where the second line implicitly defines $k_z$ as $E_{\mathbf{k}_{\parallel}}(k_z)$.
Finally, converting the expression back to discrete $k_z$ and reduced zone scheme, we obtain
\[F_{\mathbf{k}_{\parallel}}(E) = 2\pi i\sum_{n, k_z} e^{-ik_zL_z}\Theta(-v^z_{n \mathbf{k}}) \delta(E - E_{n \mathbf{k}})\].

Note that the condition $E - E_{\mathbf{k}{\parallel}}(k_z) = 0$ also introduces imaginary poles in the complex plane. Since $L_z$ is typically large, these imaginary poles cause the factor $e^{-ik_zL_z}$ to approach zero. For one-dimensional free electrons, the imaginary poles can be accounted for analytically using the dispersion relation $E{\mathbf{k}_{\parallel}}(k_z) = \hbar^2k_z^2 / 2m$. We found that the presence of imaginary poles leads to slight blurring of the delta function but does not alter its integrated value. Plugging $F_{\mathbf{k}_{\parallel}}(E)$ back into the transmission coefficient in Eq.~(\ref{eqn:mid_step}), we obtain
\begin{align*}
\tau(E) =& \frac{4\pi^2\hbar^2}{\Delta^2}\sum_{nm\mathbf{k}\mathbf{q}} |\langle \psi_{m \mathbf{q}}|\hat{\overline{j}}_z(0)|\psi_{n\mathbf{k}}\rangle |^2 e^{i(q_z - k_z)L_z}\\
&\cdot \Theta(-v^z_{m \mathbf{q}})\Theta(-v^z_{n \mathbf{k}}) \delta(E - E_{m \mathbf{q}}) \delta(E - E_{n \mathbf{k}}),
\end{align*}
where the delta functions indicate that $E_{m \mathbf{q}} = E_{n \mathbf{k}}$, consistent with the nature of "coherent"  transport. As proved in Eq.~(\ref{eqn:divergenceless}) in Appendix~\ref{app:current_density}, the current operator is divergenceless when two states have the same energy.  As a result, the conservation of current flow allows us to rewrite the position-dependent current as its average
\begin{align*}
\langle \psi_{m \mathbf{q}}|\hat{\overline{j}}_z(0)|\psi_{n\mathbf{k}}\rangle &= \frac{1}{N}\sum_z \langle \psi_{m \mathbf{q}}|\hat{\overline{j}}_z(z)|\psi_{n\mathbf{k}}\rangle\\
&= \frac{1}{N}\langle \psi_{m \mathbf{q}}|\hat{v}^z|\psi_{n\mathbf{k}}\rangle,
\end{align*}
where $N = L_z / \Delta$ is the number of grid points in the central region and $\hat{v}^z \equiv \hat{p}_z/m$ is the velocity operator. The second line is given by the property of the current density operator, as proved in Eq.~(\ref{eqn:v_paw}) in Appendix~\ref{app:paw}.
%Note that in Finite Difference framework the complete basis is given by $I = \sum_z\int dxdy |\mathbf{r}\rangle\langle\mathbf{r}|$, so it's easy to see from the definition of current density operator Eq.~(\ref{eqn:current_operator}) that the summation over averaged current density will gives velocity operator.

We note that transitions between distinct band indices ($m \neq n$) are allowed in the diffusive transport regime, leading to the same conclusion as the Kubo-Greenwood formula in the DC limit \cite{KG_formula}. However, in the coherent transport regime, certain selection rules must be followed. First, the velocity matrix element between $\mathbf{k}$ and $\mathbf{q}$ is zero. Second, we find that terms with $m \neq n$ either vanish or contribute only infinitesimally.
%The reason is explained as followed, there are two cases of degeneracy: First, the band crossing degeneracy, in this case we can neglect the contribution because the crossing region in k-space is infinitely small compared to Brillouin zone. Second case happen when the two bands are completely degenerated along some region of $k$, but it's easy to justify by degenerated perturbation theory \cite{Griffiths_QM} that in such case the off-diagonal matrix element of velocity operator
% (proportional to $\Delta H$ response of $k$ displacement) would vanish.
Thus, we can rewrite the transmission coefficient as
\begin{align*}
\tau(E) =& \frac{4\pi^2\hbar^2}{L_z^2}\sum_{n\mathbf{k}q_z} |\langle \psi_{n \mathbf{k}}|\hat{v}^z|\psi_{n\mathbf{k}}\rangle |^2\\
&\cdot \Theta(-v^z_{n \mathbf{k}}) \delta(E - E_{n \mathbf{k}}) \delta\big(E_{n, \mathbf{k}_{\parallel}}(k_z) - E_{n, \mathbf{k}_{\parallel}}(q_z)\big).
\end{align*}
Take the continuous limit for $q_z$, and the factor $\sum_{q_z} \delta\big(E_{n, \mathbf{k}_{\parallel}}(k_z) - E_{n, \mathbf{k}_{\parallel}}(q_z)\big)$ becomes
\begin{align*}
\frac{L_z}{2\pi}\int dq_z &\delta\big(E_{n, \mathbf{k}_{\parallel}}(k_z) - E_{n, \mathbf{k}_{\parallel}}(q_z)\big)\\
&= \frac{L_z}{2\pi}\int dq_z \frac{\delta(k_z - q_z)}{|dE_{n, \mathbf{k}_{\parallel}}/dq_z|} = \frac{L_z}{2\pi}\frac{1}{\hbar |v_{n \mathbf{k}}|}.
\end{align*}

Finally, we arrive at the following simplified expression for the transmission coefficient. By replacing the step function with an additional factor of $1/2$ to account for double counting and by leveraging time-reversal symmetry \cite{time_reversal}, we obtain:
\begin{equation}\label{eqn:main} \tau(E) = \frac{h}{L_z} \sum_{n, \mathbf{k}} \frac{|\langle \psi_{n \mathbf{k}} | \hat{v}^z | \psi_{n \mathbf{k}} \rangle|}{2} \delta(E - E_{n \mathbf{k}}), \end{equation}
where $\hat{v}^z $ is the velocity operator along the transport direction. This result is identical to Eq.~(\ref{eqn:semi-tau}) introduced earlier in Sec.~\ref{sec:intro}, confirming the consistency between the two derivations.

\section{Results and Discussion}\label{sec:result}

\begin{figure*}
\includegraphics[width=0.46\linewidth]{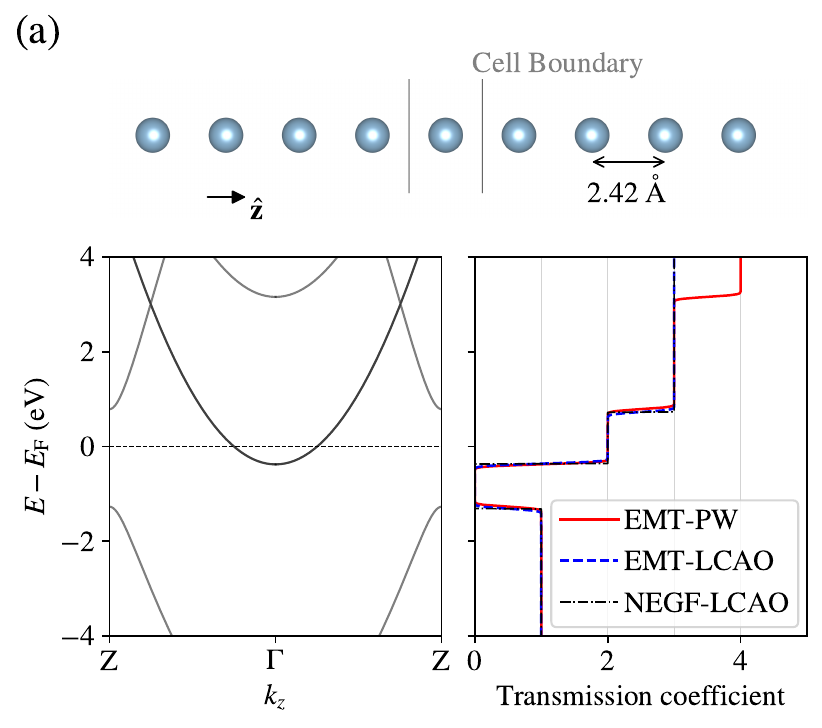}
\hspace{1em}
\includegraphics[width=0.50\linewidth]{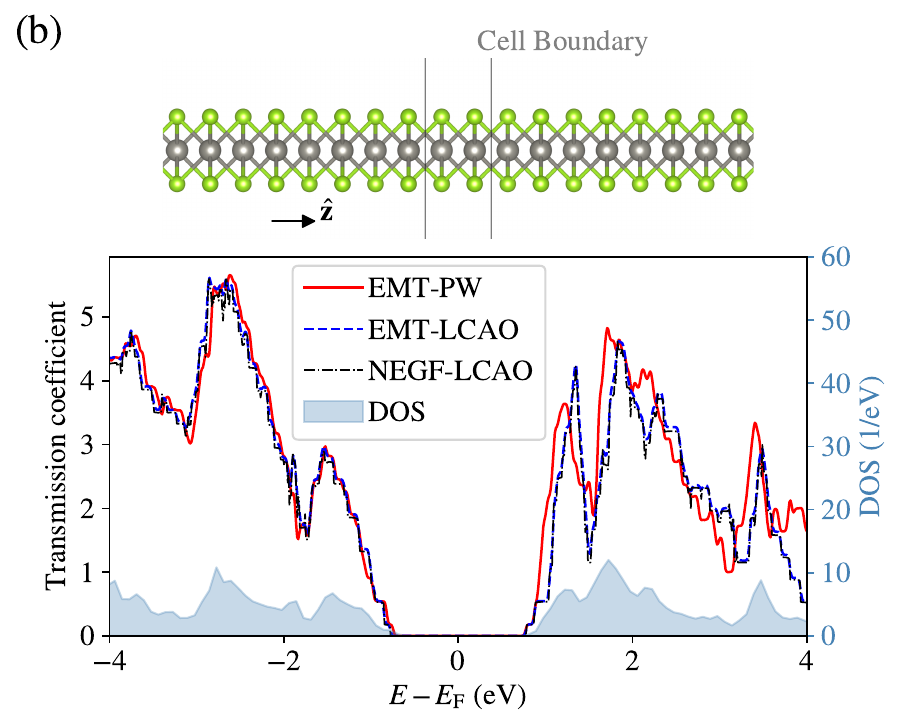}\\
\vspace{1em}
\includegraphics[width=0.46\linewidth]{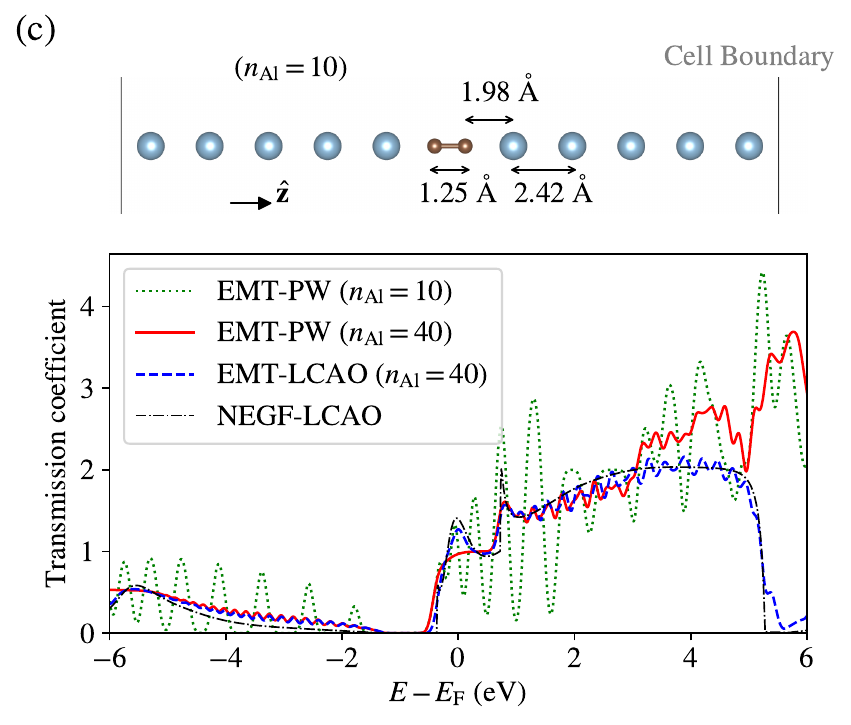}
\hspace{1em}
\includegraphics[width=0.50\linewidth]{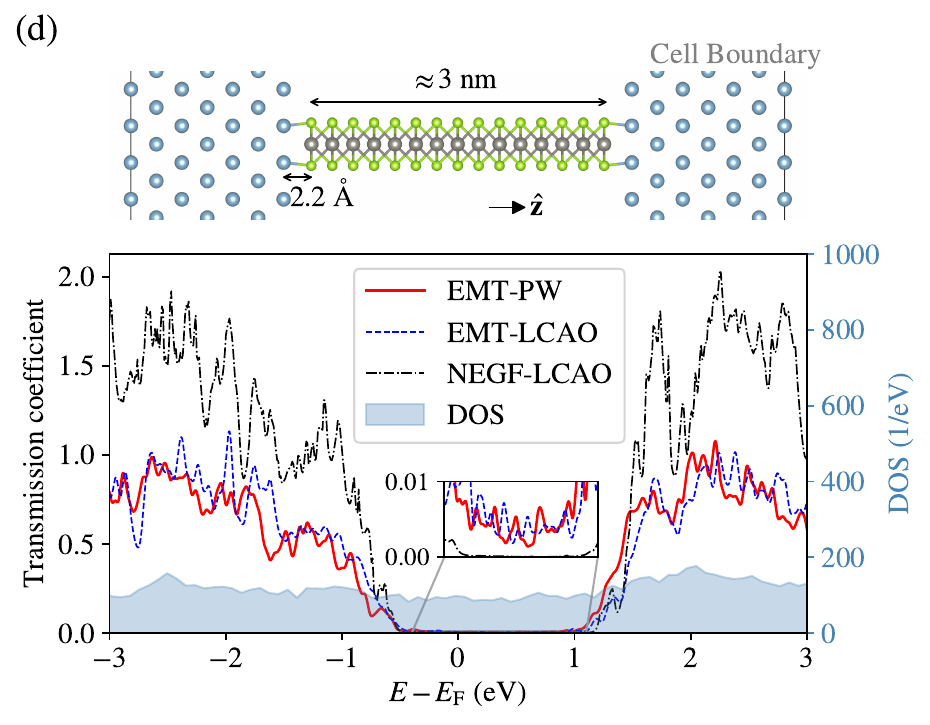}
\caption{ (color online) 
Panel (a) presents the band structure (left) and the transmission coefficient, $\tau(E)$, (right) for a one-dimensional aluminum atomic chain. Both panels use energy $E$ as the vertical axis, maintaining the same scale. The transmission coefficients $\tau(E)$, calculated using the EMT-PW (red solid line), EMT-LCAO (blue dashed line), and NEGF-LCAO (black dashed line) methods, are compared in the right panel of (a).
Similarly, $\tau(E)$ computed using the aforementioned methods is shown for (b) a two-dimensional WSe$_2$ monolayer, (c) a one-dimensional Al-C-C-Al junction, and (d) a two-dimensional Al-WSe$_2$-Al junction. The density of states (DOS) is also displayed as a blue-gray shaded area at the bottom of panels (b) and (d).
In particular, panel (c) compares $\tau(E)$ for different numbers of lead atoms ($n_{\text{Al}}=10, 40$) to verify the asymptotic behavior with respect to lead length. The reference energy is the Fermi energy $E_F$.
}
\label{fig:result}
\end{figure*}

We have developed an effective medium theory combined with density functional theory in plane-wave basos (EMT-PW), implemented within the VASP package using the Projector Augmented-Wave (PAW) method, to compute transmission coefficients. To validate the accuracy of the EMT-PW approach, we compare numerical results obtained from three different methods:
\begin{itemize}
    \item EMT-PW: Evaluation of Eq.~(\ref{eqn:main}) using a plane-wave basis with periodic boundary conditions within VASP, as described above.
    \item NEGF-LCAO: Implementation of the conventional DFT-NEGF approach using an atomic orbital basis (LCAO) via the NanoDCAL package \cite{NANODCAL1, NANODCAL2, Keldysh}.
    \item EMT-LCAO: Evaluation of Eq.~(\ref{eqn:main}) using the NanoDCAL package with velocity computed in an atomic orbital (LCAO) basis.
\end{itemize}
We focus on two types of structures, as depicted in Fig.~\ref{fig:result}:
\begin{enumerate}
    \item[(1)] Periodic systems (discussed in Subsec.~\ref{subsec:Periodic systems}):
       \begin{enumerate}
       \item[(a)] An infinitely long one-dimensional aluminum atomic chain (1D).
       \item[(b)] A two-dimensional WSe$_2$ monolayer (2D).     
       \end{enumerate}
    \item[(2)] Junction systems (discussed in Subsec.~\ref{subsec:Junction systems}): 
       \begin{enumerate}
       \item[(c)] A one-dimensional Al–C–C–Al atomic junction.
       \item[(d)] A two-dimensional Al–WSe$_2$-Al heterojunction.     
       \end{enumerate}
\end{enumerate}

We provide a brief overview of the methods in Subsec.~\ref{subsec:Methods} and present the results for periodic systems in Subsec.~\ref{subsec:Periodic systems} and for junction systems in Subsec.~\ref{subsec:Junction systems}. In Subsec.~\ref{subsec:effective-gate}, we demonstrate that combining EMT-PW with effective gate theory offers a convenient approach for studying quantum transport properties in nanojunctions under the application of a gate voltage.

\subsection{Methods}\label{subsec:Methods}

The EMT-PW approach calculates transmission coefficients based on Eq.~(\ref{eqn:main}), where the velocity operator is evaluated numerically using a plane-wave (PW) basis within VASP \cite{vasp1, vasp2, vasp3}. We employ PAW pseudopotentials \cite{uspp_to_paw} and the Perdew-Burke-Ernzerhof (PBE) exchange-correlation functional \cite{PBE}. The Dirac delta function in Eq.~(\ref{eqn:main}) is numerically approximated using a normalized Gaussian function with an appropriate broadening width.

A central challenge in this methodology is accurately evaluating the expectation value of the velocity operator within the PAW formalism. In the PAW method, the all-electron wavefunction $\Psi_{n\mathbf{k}}$
is reconstructed from the pseudo-wavefunction $\widetilde{\Psi}_{n\mathbf{k}}$ via a linear transformation \cite{paw, optics_transverse}. To address this, we employ a more robust approach by using the following relation: $\langle \psi_{n \mathbf{k}}|\hat{v}^z |\psi_{n\mathbf{k}}\rangle = (1/\hbar)\partial E_{n \mathbf{k}} / \partial k_z$.

The detailed derivation of $\partial E_{n \mathbf{k}} / \partial k_z$ within the PAW formalism is provided in Appendix~\ref{app:paw}, culminating in Eq.~(\ref{eqn:v_paw}). This method has been successfully implemented in VASP by Gajdoš for the calculation of optical matrix elements \cite{optics_longitudinal}. We applied these computational frameworks to the systems illustrated in Fig.~\ref{fig:result}. Detailed parameters for the VASP and NanoDCAL simulations, including k-point sampling, energy cutoffs, and convergence criteria, are provided in Appendix~\ref{app:para}. 

\subsection{Periodic systems}\label{subsec:Periodic systems}

The left panel of Fig.~\ref{fig:result}(a) shows the band structure $E_{nk}$, calculated using VASP, for the aluminum atomic chain, with the Fermi energy, $E_F$, set as the reference energy. The band structure reveals that the electron eigenstates behave similarly to those of a nearly free-electron gas. Aside from band degeneracy and band folding, the energy bands are well described by parabolic dispersion curves.

The energy $E$ defines a horizontal plane that intersects the energy bands, forming equi-energy surfaces in $\mathbf{k}$-space. In the one-dimensional Al chain, the energy $E$ intersects the $n-$th energy band, $E_{nk_z}$, at two points corresponding to the wavevectors $k^{\pm}_{n}(E)=\pm \sqrt{2m(E-E_{nC})/\hbar^2}$.  Note that only $k^{+}_{n}$, which belongs to $k_L \equiv \{\mathbf{k} | \mathbf{v}_z > 0\}$ and corresponds to the electron incident from the left, contributes to the transmission coefficient $\tau(E)$ via the component $\tau_n(E)$. As shown in Appendix~\ref{app:taunEeq1}, the relation $\tau_n(E) = T_{nk^{+}_{n}(E)}  \cdot D_{nk_L}(E)$ holds, where the transmission probability density $T_{nk}$ at $k=k^{+}_{n}(E)$ is given by $T_{nk_{n}^{+}(E)} = \frac{h}{L_z}\frac{\hbar k_{n}^{+}(E)}{m}$ and the partial density of states (PDOS) at $k=k^{+}_{n}(E)$ is $D_{nk_L}(E) =\frac{mL_z}{2\pi \hbar^2 k_{n}^{+}(E)}$. As a result, $\tau_n(E) = 1$.

The right panel of Fig.~\ref{fig:result}(a) shows the total transmission coefficient, $\tau(E)$, which aggregates contributions from all eigenchannels. When the horizontal line defined by $E$ intersects the energy bands $E_{nk}$, each intersected band increases $\tau(E)$ by one. Consequently, the transmission coefficient $\tau(E)$ takes integer values, determined by the number of energy bands participating in $\tau(E) = \sum_{n} \tau_n(E)$. An exception occurs at $E = E_F$, where the band intersected by the Fermi energy contributes 2 to $\tau(E)$. This is due to the two-fold degeneracy of that energy band.

Figure~\ref{fig:result}(b) shows the transmission coefficients calculated using the EMT-PW, EMT-LCAO, and NEGF-LCAO methods for a two-dimensional WSe$_2$ monolayer. The density of states (DOS) is also displayed at the bottom. All three methods consistently predict the same band gap for this 2D semiconductor.

The transmission coefficients, $\tau(E)$, obtained from the three methods are in good agreement, except in the high-energy region. The discrepancies observed at higher energies are likely due to the limitations of the finite LCAO basis set used in NEGF-LCAO and EMT-LCAO, which fail to accurately reproduce certain high-energy bands.

\subsection{Junction systems}\label{subsec:Junction systems}

It should be noted that NEGF-LCAO (NanoDCAL) operates within an open system framework, incorporating leads into the calculation by matching boundary conditions. In contrast, EMT-PW (VASP) is applicable only to periodic systems, where the leads are simulated within a supercell. The effectiveness of using EMT-PW for junction systems has yet to be fully explored.

To evaluate the accuracy of the EMT-PW method, we compute $\tau(E)$ for a one-dimensional junction system consisting of two aluminum chains as leads, with a carbon-carbon bond in the middle acting as the scattering region \cite{Ni-C-Ni_chain}, as illustrated at the top of Fig.~\ref{fig:result}(c). Ideally, the aluminum chains serving as leads should be infinitely long to ensure that the EMT-PW framework accurately captures the behavior of infinite leads, similar to how NEGF-LCAO models open systems. To investigate this, we examine the asymptotic behavior in EMT-PW by increasing the lead lengths, adjusting the number of aluminum atoms accordingly.

First, we compare $\tau(E)$ for Al-C-C-Al junctions computed using EMT-PW with electrodes containing 10 and 40 Al atoms, as shown in Fig.~\ref{fig:result}(c). For the shorter lead length with ten Al atoms [denoted as $\text{EMT-PW} (n_{\text{Al}} = 10)$], a periodic oscillatory behavior in $\tau(E)$ is observed, resulting from the quantization imposed by the periodic boundary conditions of the superlattice. For the longer lead length with 40 Al atoms [denoted as $\text{EMT-PW}(n_{\text{Al}} = 40$)], the oscillatory behavior is significantly suppressed and smoothed out using Gaussian smearing.

Second, the result for the longer lead length with 40 Al atoms using EMT-PW [denoted as $\text{EMT-PW}(n_{\text{Al}} = 40)$] is compared with those computed using EMT-LCAO with 40 atoms in the lead [denoted as $\text{EMT-LCAO}(n_{\text{Al}} = 40)$], as well as NEGF-LCAO. The $\tau(E)$ for $\text{EMT-LCAO}(n_{\text{Al}} = 40)$ closely matches the NEGF-LCAO calculations. Both results are consistent with that obtained from $\text{EMT-PW}(n_{\text{Al}} =40)$, though some discrepancies appear in the high-energy region ($E > 5$~eV). These discrepancies are attributed to the limitations of the finite LCAO basis set, which fails to capture high-energy bands, similar to the behavior observed in Fig.~\ref{fig:result}(b) discussed in Subsec.~\ref{subsec:Periodic systems}.

In addition,  quantum unit transmission coefficient is observed in [$\text{EMT-PW}(n_{\text{Al}} = 40)$] using VASP simulation package while such quantization phenomenon are missed in both cases of  $\text{EMT-LCAO}(n_{\text{Al}} = 40)$ and NEGF-LCAO using NanoDCAL. At around $E=E_F$, $\tau(E)$ exhibiting a plateau with $\tau(E) \approx 1$ in $\text{EMT-PW}(n_{\text{Al}} = 40)$, while $\tau(E)$ exhibits a peak instead. 

%The ability of EMT-PW to present quantized plateaus near $E_F$ can be attributed to two key factors: (1) VASP employs a larger plane-wave basis set compared to the finite atomic orbital basis used in NanoDCAL, and (2) VASP utilizes a more accurate PAW pseudopotential. This comparison highlights a valuable aspect of our EMT-PW approach. (replaced by the folowing paragraph in the resubmission version)

The disagreement between EMT-PW and NEGF-LCAO primarily stems from differences in the band structures computed using the plane-wave (PW) and linear combination of atomic orbital (LCAO) basis sets, especially near the Fermi energy. Since the plane-wave basis is a complete set and generally more reliable than LCAO, we consider the quantized plateaus observed near $E_F$ in our EMT-PW results to be both robust and physically meaningful. This comparison highlights a key strength of the EMT-PW approach in accurately capturing essential features of quantum transport that are closely tied to the underlying band structure.

In conventional NEGF calculations, atomic basis sets are typically used due to computational limitations. However, with the EMT-PW framework, we can perform transmission coefficient calculations using a plane-wave basis set, enabling more precise results—such as the quantized plateau observed in our Al-C-C-Al system.

Fig.~\ref{fig:result}(d) compares $\tau(E)$ computed using EMT-PW, EMT-LCAO, and NEGF-LCAO for the Al-WSe$_2$-Al junction. The junction consists of a two-dimensional semiconducting WSe$_2$ monolayer, approximately 3 nm in length, sandwiched between two bulk aluminum electrodes. Unlike the Al-C-C-Al junction with one-dimensional electrodes, we find that the transmission coefficient calculated from EMT-PW is relatively unaffected by the lead length.

The results of both EMT-PW and EMT-LCAO, which use periodic boundary conditions, exhibit a band gap that aligns well with the band gap computed from NEGF-LCAO, which models an open system. All three methods reveal a transmission gap and display P-type semiconducting tunneling junction behavior.

However, a larger discrepancy in $\tau(E)$ outside the band gap is observed between the results from EMT-PW and EMT-LCAO (both using periodic boundary conditions) and NEGF-LCAO (which models an open system with leads extended to infinity through boundary condition matching). The methods with periodic boundary conditions tend to yield smaller $\tau(E)$ values compared to NEGF-LCAO.

Despite these differences, $\tau(E)$ calculated from the three methods qualitatively agrees. Note that while in Fig.\ref{fig:result}(b), the transmission drops to zero due to the absence of states within the energy gap, in Fig.\ref{fig:result}(d), a transmission gap exists even when the density of states remains non-zero in that region.

\subsection{EMT-PW combined with effective gate odel}\label{subsec:effective-gate}

%{add the followings for resubmission 05/12/2025}

According to numerical calculations using Nanodcal, the application of a gate voltage $V_g$ causes a change in the chemical potentials $\mu_L$ and $\mu_R$, initially at $V_g = 0$, with an energy approximately proportional to $V_g$. When the chemical potentials lie within the band gap, they shift by about 83\% of $eV_g$. However, as the system becomes more conductive, i.e., when the chemical potentials lie outside the band gap, the gate voltage becomes less effective at shifting the chemical potentials. Numerical calculations using NEGF-DFT show that in such conductive regimes, the chemical potentials shift by only about 33\% of $eV_g$. Based on these observations, we construct an empirical model, referred to as the \emph{effective gate model}, in which the chemical potentials are effectively shifted by $eV_{\text{G}}^{\text{eff}}(V_g)$, as illustrated in Ref.~\cite{FET-AlN}. For the EMT-PW method, the Landauer formula combined with the effective gate model can be applied to study the current-voltage characteristics of nanojunctions under an applied gate voltage $V_g$. 

When a gate voltage $V_g$ is applied, the chemical potentials shift by an energy of $eV_{\text{G}}^{\text{eff}}(V_g)$, where $V_{\text{G}}^{\text{eff}}(V_g)$ is the effective gate voltage. The effective gate potential is characterized by two parameters, $\alpha_{\text{in}}$ and $\alpha_{\text{out}}$, which empirically describe the gate-controlling efficiency for the nanojunction under specific gate artitecture. These parameters determine how $V_{\text{G}}^{\text{eff}}(V_g)$ shifts the chemical potential upon the applied $V_g$. By combining the transmission coefficient $\tau(E)$, computed using the EMT-PW method, with the effective gate model, we analyze a field-effect transistor (FET) formed by the Al-WSe$_2$-Al junction, using a gate architecture similar to that described in Ref.~\cite{FET-AlN}.

\begin{figure} 
\includegraphics[width=\linewidth]{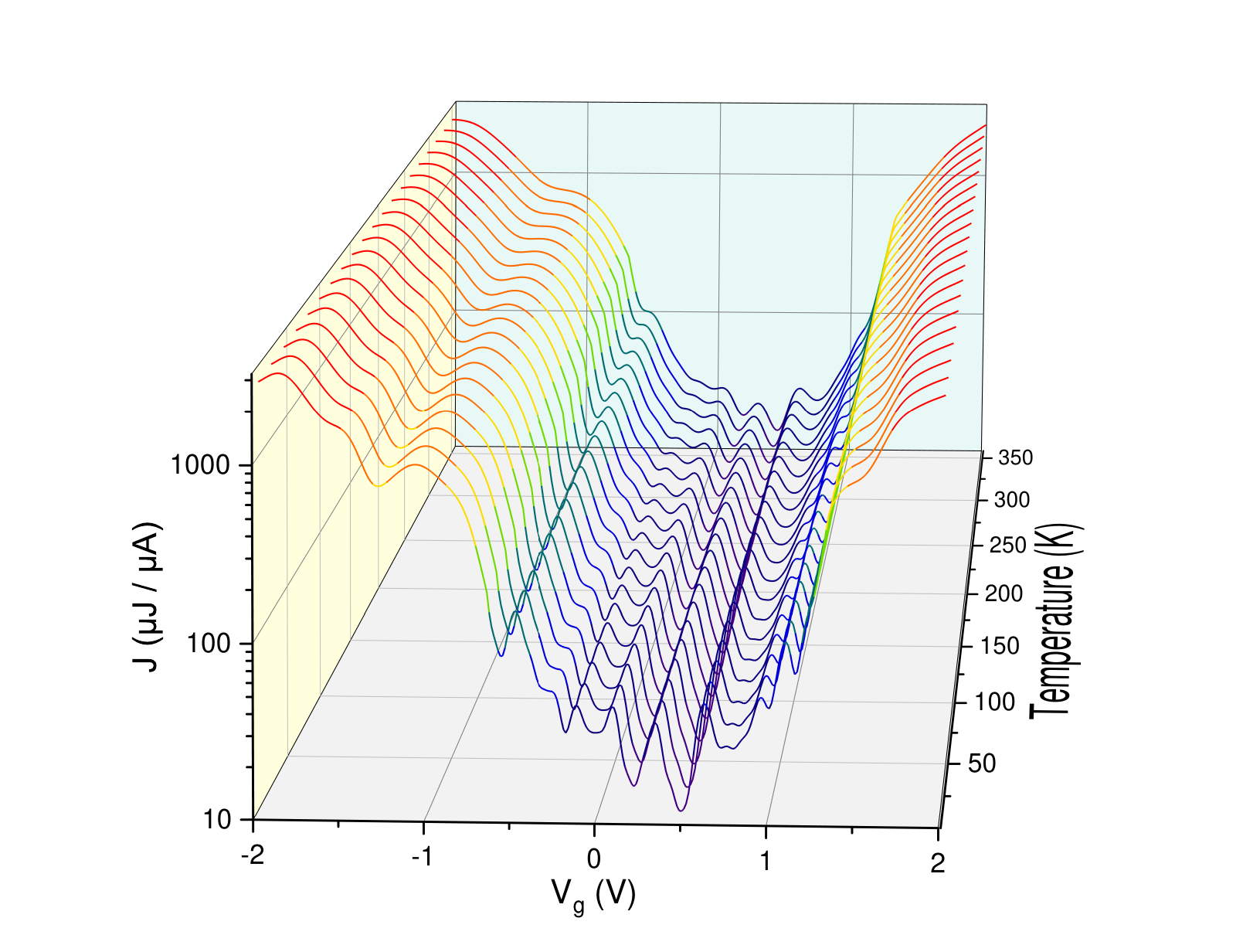}
\caption{ (color online)
The current density of the Al-WSe$_2$-Al junction as a function of the gate voltage V$_g$ and temperature $T$ at $\text{V}_{\text{ds}}=50~$mV. The results are determined by utilizing $\tau(E)$ from the EMT-PW theory and $V_{\text{G}}^{\text{eff}}$ from the effective gate mode.
}
\label{fig:current}
\end{figure}

Figure~\ref{fig:current} shows the current density as a function of the applied gate voltage $V_g$ and temperature, calculated using the combination of EMT-PW and the effective gate model. The gate-controlling efficiency parameters are set to $\alpha_{\text{in}} = 83$~\% and $\alpha_{\text{out}} = 33$~\%, corresponding to whether the chemical potential lies inside or outside the band gap, respectively. This approach enables efficient calculation of the current density as a function of $V_g$ and temperature $T$, as shown in Fig.~\ref{fig:current}.

It is important to note that the error in the effective gate model is controllable. We have demonstrated that the error is within $[ {\ln(10)}/{S.S.}]  | \Delta V_{\text{G}}^{\text{eff}} |$, $S.S.$ is the subthreshold swing in Ref.~\cite{FET-AlN}.

\section{Conclusion}

Using VASP with the projector augmented-wave (PAW) technique and a plane-wave basis set, we developed an effective medium theory (EMT-PW) capable of calculating the transmission coefficient $\tau(E)$ from first-principles approaches. We derive $\tau(E)$ in EMT-PW in three ways from $J=nqv$, the field operator, and the NEGF transmission coefficient $\tau = \text{Tr}[\Gamma_L G_r \Gamma_R G_a]$. We demonstrate that $\tau(E)$ can be computed from $v^z_{n\mathbf{k}}$ [Eq.(\ref{eqn:v_paw})], derived from $J = nqv$, or from $\langle \psi_{n\mathbf{k}}(\mathbf{r}) | \frac{\hat{p}_z}{m} | \psi_{n\mathbf{k}}(\mathbf{r}) \rangle$ [Eqs.(\ref{eq:TnK}) and (\ref{eqn:tau(E)1})], derived from the field operator.

The transmission coefficient $\tau(E)$ computed using EMT-PW is compared with results from NanoDCAL, which employs an atomic orbital basis set under either non-periodic (NEGF-LCAO) or periodic (EMT-LCAO) boundary conditions. To asymptotically approach the NEGF-DFT results for junction systems, EMT-PW requires an increased number of metal atoms in the electrode regions. The comparison confirms the validity of our EMT-PW theory, showing general agreement with minor discrepancies attributed to differences in basis sets, pseudopotentials, and the implementation of periodic versus non-periodic boundary conditions. Notably, discrepancies in $\tau(E)$ in the high-energy region arise due to the finite basis set used in NanoDCAL, which omits contributions from high-energy bands. The adoption of PAW pseudopotentials in the EMT-PW method enhances the accuracy of $\tau(E)$.

By partitioning the wavefunctions {$\Psi_{n\mathbf{k}}(\mathbf{r})$}, which form a complete set, into two subsets, the EMT-PW method effectively circumvents the common issue "overcompleteness" of LCAO basis in non-equilibrium transport theory. Moreover, EMT-PW allows for the decomposition of $\tau(E)$ into contributions from individual eigenstates: $\tau(E) = \sum_n \tau_n(E)$, where $\tau_n(E) = \sum_{\mathbf{k}_L} T_{n\mathbf{k}_L} \delta(E - E_{n\mathbf{k}_L})$, and $T_{n\mathbf{k}_L}=\frac{\langle \psi_{n\mathbf{k}_{L}}(\mathbf{r}) | \hat{p}_z | 
\psi_{n{\mathbf{k}}_{L} }(\mathbf{r}) \rangle}{m}$.
\deleted[comment={delete}]{Here, the transmission probability density $T_{n\mathbf{k}_L}$ is defined for the $n$-th energy band $E_{n\mathbf{k}}$ and weighted by the PDOS $D_{n\mathbf{k}_L}(E)$. This framework enables the effective study of quantum transport characteristics in gate-controlled nanodevices when combined with the effective-gate model.}

\deleted[comment={delete}]{The one-electron wavefunction $\psi_{n\mathbf{k}}(\mathbf{r})$ forms a complete basis set for the EMT-PW theory, providing a robust foundation for exploring the quantum statistics of electrons and the quantum correlations of currents within the second quantization framework. EMT-PW surpasses NEGF-DFT in its capability to investigate quantum many-body states in second quantization, offering a powerful tool for advanced quantum transport studies.}

In essence, our work establishes a robust framework for exploring 
\replaced{equilibrium quantum transport properties}{quantum statistics
of electrons and current correlations} 
using second-quatized form. 
\replaced{Without using Green's function, the}{The} 
field operator constructed from the one-electron wavefunction within a complete set will provide
\replaced{an alternative tool to extract transport current, quantum statistics, and many-body effect.}{a
valuable tool for examining many-body effects in nanoscale systems. This framework
provides a solid theoretical foundation to understanding quantum behavior resulting
from the discrete nature of electrons, paving the way for advances in nanoscale tech-
nology and quantum transport research.}

\appendix

\section{Current Density Operator}\label{app:current_density}
In quantum mechanics, the probability current density is a function of position, represented as
\[\mathbf{j}(\mathbf{r}) = \frac{\hbar}{2mi}\big(\psi^*(\mathbf{r}) \nabla \psi(\mathbf{r}) - \psi(\mathbf{r}) \nabla \psi^*(\mathbf{r})\big),\]
from which we can define the probability current density operator as,
\begin{equation}\label{eqn:current_operator_a}
\hat{\mathbf{j}}(\mathbf{r}) = \frac{1}{2m}\big(\ \hat{\mathbf{p}}|\mathbf{r}\rangle\langle\mathbf{r}| + |\mathbf{r}\rangle\langle\mathbf{r}|\hat{\mathbf{p}}\ \big)
\end{equation}
To see this, we can consider its matrix element,
\begin{align*}
\mathbf{j}_{mn}(\mathbf{r}) &= \langle\psi_m|\hat{\mathbf{j}}(\mathbf{r})|\psi_n\rangle \\&= 
\frac{1}{2m}\big(\langle\psi_m|\hat{\mathbf{p}}|\mathbf{r}\rangle\langle\mathbf{r}|\psi_n\rangle + \langle\psi_m|\mathbf{r}\rangle\langle\mathbf{r}|\hat{\mathbf{p}}|\psi_n\rangle\big)\\
&=\frac{1}{2m}\big((i\hbar\nabla\langle\psi_m|\mathbf{r}\rangle)\langle\mathbf{r}|\psi_n\rangle + \langle\psi_m|\mathbf{r}\rangle(-i\hbar\nabla\langle\mathbf{r}|\psi_n\rangle)\big)\\
&=\frac{\hbar}{2mi}\big(\psi_m^*(\mathbf{r}) \nabla \psi_n(\mathbf{r}) - \psi_n(\mathbf{r}) \nabla \psi_m^*(\mathbf{r})\big),
\end{align*}
where we have used $\langle\mathbf{r}|\hat{\mathbf{p}}|\psi\rangle = -i\hbar\nabla\psi(\mathbf{r})$ \cite{Sakurai} and it's complex conjugate. Setting $m=n$ simplifies the expression to the original definition of current density, $\mathbf{j}(\mathbf{r})$.

An important property of the probability current is that it satisfies the continuity equation, which simplifies to $\nabla \cdot \mathbf{j} = 0$ for a stationary state. However, what happens if the current operator is sandwiched between two different states? To address this, we must refer to the Schrödinger equation to find the answer:
\begin{align*}
&-\frac{\hbar^2}{2m}\nabla^2 \psi_n(\mathbf{r}) + V(\mathbf{r}) \psi_n(\mathbf{r}) = E_n \psi_n(\mathbf{r})\\
&-\frac{\hbar^2}{2m}\nabla^2 \psi^*_m(\mathbf{r}) + V(\mathbf{r}) \psi^*_m(\mathbf{r}) = E_m \psi^*_m(\mathbf{r}).
\end{align*}
We multiply the preceding equations by $\psi^*_m$ and $\psi_n$ respectively, and subtract the two. We obtain
\begin{align*}
&(E_n - E_m)\psi_n(\mathbf{r})\psi^*_m(\mathbf{r}) \\
&=-\frac{\hbar^2}{2m}\big[\psi^*_m(\mathbf{r})\nabla^2 \psi_n(\mathbf{r}) - \psi_n(\mathbf{r})\nabla^2 \psi^*_m(\mathbf{r})\big]\\
&=-\frac{\hbar^2}{2m}\nabla\cdot\big[\psi^*_m(\mathbf{r})\nabla \psi_n(\mathbf{r}) - \psi_n(\mathbf{r})\nabla \psi^*_m(\mathbf{r})\big]\propto \nabla\cdot\mathbf{j}_{mn}(\mathbf{r}).
\end{align*}
It is now clear that this matrix element is not divergence-free unless $E_m = E_n$. Thus, we might conclude that
\begin{equation}\label{eqn:divergenceless}
\nabla\cdot\mathbf{j}_{mn}(\mathbf{r}) = 0 \quad\text{if}\quad E_m = E_n.
\end{equation}

Another useful relation can be derived from the completeness relation within the finite-difference framework. $1 = \sum_z |z\rangle\langle z|$, accordingly we have,
\[1 = \sum_z \int \text{d}x\text{d}y|\mathbf{r}\rangle\langle \mathbf{r}|,\]
Plug into the definition of current density operator Eq.~(\ref{eqn:current_operator_a}), we find that the summation (integral) of current density operator gives velocity operator
\begin{equation}\label{eqn:current_velocity}
\sum_z \int \text{d}x\text{d}y\ \hat{\mathbf{j}}(\mathbf{r}) = \frac{\hat{\mathbf{p}}}{m} = \hat{\mathbf{v}}.
\end{equation}
\section{Velocity in PAW method}\label{app:paw}
To obtain the expectation value of the velocity operator using VASP in PAW method, we begin with the Kohn-Sham equation in the PAW formalism, \cite{uspp_to_paw}
\[\hat{H}|\tilde{\psi}_{n \mathbf{k}}\rangle = E_{n \mathbf{k}} \hat{S}|\tilde{\psi}_{n \mathbf{k}}\rangle ,\]
where 
\begin{equation}\label{eqn:H_paw}
\hat{H} = -\frac{\hbar^2}{2m}\nabla^2 + \tilde{V}_{\text{eff}} + \sum_{ij}|\tilde{p}_i\rangle D_{ij} \langle \tilde{p}_j | ,
\end{equation}
and 
\begin{equation}\label{eqn:S_paw}
\hat{S} = 1 + \sum_{ij}|\tilde{p}_i\rangle Q_{ij} \langle \tilde{p}_j |.
\end{equation}
where $|\tilde{p}_i\rangle$ are the projectors that account for the non-local contribution in PAW formalism. Note that the basis of VASP in PAW method is non-orthogonal, and satisfies $\langle\tilde{\Psi}_m|\hat{S}|\tilde{\Psi}_n\rangle = \delta_{mn}$. 

Notably, the wavefunction calculated using the PAW Hamiltonian differs from the all-electron wavefunction under the bare Coulomb potential, even though the eigenenergy should theoretically be the same. As a result, the most accurate way to compute the expectation value of velocity which is $v^z_{n \mathbf{k}} = (1/\hbar)\partial E_{n \mathbf{k}} / \partial k_z$.
%instead of $\langle \psi_{n \mathbf{k}} |\hat{p}_z| \psi_{n \mathbf{k}}\rangle / m$ since the "state" is less trustable.

To obtain the formula for the derivatives of eigen-energy, we need to utilize first order perturbation to find the response of $E_{n \mathbf{k}}$ with respect to change of $\mathbf{k}$. First, rewrite the Kohn-Sham equation as, a periodic function.
\[\hat{H}_{\mathbf{k}}|\tilde{u}_{n \mathbf{k}}\rangle = E_{n \mathbf{k}} \hat{S}_{\mathbf{k}}|\tilde{u}_{n \mathbf{k}}\rangle ,\]
where we define the k-dependent Hamiltonian $\hat{H}_{\mathbf{k}} = e^{-i\mathbf{k}\cdot\mathbf{r}} \hat{H} e^{i\mathbf{k}\cdot\mathbf{r}}$ and $\hat{S}_{\mathbf{k}} = e^{-i\mathbf{k}\cdot\mathbf{r}} \hat{S} e^{i\mathbf{k}\cdot\mathbf{r}}$. Next, we consider a small change in $\mathbf{k}\rightarrow \mathbf{k}+\delta \mathbf{k}$, and we obtain,
\begin{equation*}
\begin{aligned}
& (\hat{H}_{\mathbf{k}} + \delta \hat{H}_{\mathbf{k}})(|\tilde{u}_{n \mathbf{k}}\rangle + | \delta\tilde{u}_{n \mathbf{k}}\rangle) \\
&= (E_{n \mathbf{k}} + \delta E_{n \mathbf{k}}) (\hat{S}_{\mathbf{k}} + \delta \hat{S}_{\mathbf{k}})(|\tilde{u}_{n \mathbf{k}}\rangle + | \delta\tilde{u}_{n \mathbf{k}}\rangle).   
\end{aligned}
\end{equation*}
Collect the first order terms from the above equation
\begin{equation*}
\begin{aligned}
& \delta \hat{H}_{\mathbf{k}}|\tilde{u}_{n \mathbf{k}}\rangle + \hat{H}_{\mathbf{k}} | \delta\tilde{u}_{n \mathbf{k}}\rangle  \\
&= \delta E_{n \mathbf{k}} \hat{S}_{\mathbf{k}}|\tilde{u}_{n \mathbf{k}}\rangle + E_{n \mathbf{k}} \delta\hat{S}_{\mathbf{k}}|\tilde{u}_{n \mathbf{k}}\rangle + E_{n \mathbf{k}} \hat{S}_{\mathbf{k}}|\delta\tilde{u}_{n \mathbf{k}}\rangle.
\end{aligned}
\end{equation*}
Sandwiching $\langle \tilde{u}_{n \mathbf{k}}|$ on both sides of the equation, it's simple to see the terms involve $|\delta\tilde{u}_{n \mathbf{k}}\rangle$ canceled, and by $\delta E_{n \mathbf{k}} = \delta \mathbf{k} \cdot (\partial E_{n \mathbf{k}} / \partial \mathbf{k})$, we have
\[\frac{\partial E_{n \mathbf{k}}}{\partial \mathbf{k}} = \langle \tilde{u}_{n \mathbf{k}}|\left(\frac{\partial \hat{H}_{\mathbf{k}}}{\partial \mathbf{k}} - E_{n \mathbf{k}}\frac{\partial \hat{S}_{\mathbf{k}}}{\partial \mathbf{k}}\right)|\tilde{u}_{n \mathbf{k}}\rangle,\]
which is the non-orthogonal version of Hellmann-Feynman theorem. Using the definition of $\hat{H}_{\mathbf{k}} = e^{-i\mathbf{k}\cdot\mathbf{r}} \hat{H} e^{i\mathbf{k}\cdot\mathbf{r}}$, we get $\partial H_{\mathbf{k}}/\partial \mathbf{k} = i[H_{\mathbf{k}}, \mathbf{r}]$. So, we can rewrite
\[\mathbf{v}_{n \mathbf{k}} = \frac{1}{\hbar}\frac{\partial E_{n \mathbf{k}}}{\partial \mathbf{k}} = \frac{i}{\hbar}\langle\tilde{\psi}_{n \mathbf{k}}| [\hat{H} - E_{n \mathbf{k}}\hat{S}, \mathbf{r}] |\tilde{\psi}_{n \mathbf{k}}\rangle\]
Finally, by substituting the definition in Eq.~(\ref{eqn:H_paw}) for $\hat{H}$ and Eq.~(\ref{eqn:S_paw}) for $\hat{S}$, we obtain,
\begin{equation}\label{eqn:v_paw}
\begin{aligned}
\mathbf{v}_{n \mathbf{k}} & = \frac{1}{m} \langle\tilde{\psi}_{n \mathbf{k}}| (-i\hbar \nabla) |\tilde{\psi}_{n \mathbf{k}}\rangle \\
& + \frac{i}{\hbar}\sum_{ij} \langle \tilde{\psi}_{n \mathbf{k}}|\tilde{p}_i\rangle (D_{ij} 
- E_{n \mathbf{k}} Q_{ij}) \langle \tilde{p}_j |\mathbf{r}|\tilde{\psi}_{n \mathbf{k}}\rangle \\
&\qquad - \langle \tilde{\psi}_{n \mathbf{k}} |\mathbf{r}|\tilde{p}_i\rangle (D_{ij} - E_{n \mathbf{k}} Q_{ij}) \langle \tilde{p}_j|\tilde{\psi}_{n \mathbf{k}}\rangle.
\end{aligned}
\end{equation}

\section{Parameters in Calculations}\label{app:para}

The Al-C-C-Al junction is relaxed using the Vienna Ab-initio Simulation Package (VASP). In the Al-WSe\_2-Al system, the bulk structures of Al and WSe$_2$ remain fixed while the bond length is optimized by minimizing the total energy. The calculation of the EMT-PW method is implemented with VASP. The evaluation of Eq.~(\ref{eqn:main}) requires Gaussian smearing, which approximates the delta-function, $\delta[E-E_n(\mathbf{k})]$, by a gaussian function, i.e.,
\[\delta(E - E_{n\mathbf{k}}) \approx \frac{1}{\sqrt{\pi} \sigma} e^{-\frac{(E - E_{n\mathbf{k}})^2}{\sigma^2}},\]
where $\sigma$ is the standard deviation. Table~\ref{tb:vasp} lists the key parameters utilized in VASP and $\sigma$.
\begin{table}[h]
\centering
\caption{
Key parameters in EMT-PW (VASP)
}
\begin{ruledtabular}
\begin{tabular}{lcccc}
Structure          & ENCUT  & $k-$pointS                       & EDIFF        & $\sigma$   \\
                   & (eV)  &                        & (eV)         &  (eV)  \\
\hline
Al chain           & 600         & 1 1 501                       & $1\times 10^{-5}$  & 0.05 \\
WSe$_2$ monolayer  & 223.1       & 11 1 501                      & $1\times 10^{-5}$  & 0.02 \\
Al-C-C-Al          & 600         & 1 1 51\footnote{For $n_{\text{Al}} = 40$ calculation.} & $1\times 10^{-5}$  & 0.1   \\
Al-WSe$_2$-Al      & 400         & 11 3 11                       & $1\times 10^{-6}$  & 0.03  \\
\end{tabular}
\end{ruledtabular}
\label{tb:vasp}
\end{table}

The calculation of the NEGF-LCAO method is implemented with NanoDCAL. The $k-$points of the scattering region are identical to the k points of the lead, with the exception that the number of $k-$points in the z-direction is one. The electronic iteration stops when both ‘hMatrix’ and ‘rhoMatrix’ reach its convergence criteria. Table~\ref{tb:NanoDCAL} lists the key parameters chosen for use in the NanoDCAL.
\begin{table}[h]
\centering
\caption{
Key parameters in NEGF-LCAO (NanoDCAL)
}
\begin{ruledtabular}
\begin{tabular}{lccc}
Structure & Energy cutoff  & $k-$points   & convergence criteria \\
 &  (in Hartree) &  (lead)  &  (in Hartree) \\

\hline
Al chain      & 80  & 1 1 100  & $1\times 10^{-5}$ \\ 
WSe$_2$       & 80  & 11 1 100 & $1\times 10^{-5}$ \\ 
Al-C-C-Al     & 80  & 1 1 100  & $1\times 10^{-5}$ \\ 
Al-WSe$_2$-Al & 100 & 11 3 100 & $1\times 10^{-5}$ \\ 
\end{tabular}
\end{ruledtabular}
\label{tb:NanoDCAL}
\end{table}

Finally, the EMT-LCAO calculations use the same energy cutoff and convergence criteria as the NEGF-LCAO method listed in Table~\ref{tb:NanoDCAL}, while the $k$-points and $\sigma$ are chosen to match those used in the EMT-PW calculations shown in Table~\ref{tb:vasp}.

\section{$\tau_n(E)=1$ for 1D parabolic bands}\label{app:taunEeq1}

For the aluminum atomic chain, the one-dimensional crystal band structure consists of energy bands that closely resemble those of a free-electron gas, as shown in the left panel of Fig.~\ref{fig:result}(a). In this case, the transmission coefficient, $\tau_n(E)$, associated with each crystal band $E_{nk}$, is equal to one. To demonstrate this, we assume that each band follows a quadratic dispersion relation, with its minimum energy of the band set as $E_{nC}$:
\begin{equation}
    E_{nk}=E_{nC}+\frac{\hbar^2k^2}{2m},
    \label{eqn:free-electron-band}
\end{equation}
where $n$ is the band index. Given this one-dimensional parabolic energy dispersion, the transmission coefficient for the $n$-th energy band is expressed as:
\begin{equation}
    \tau_{n}=\sum_{k_L} T_{nk_L} \delta(E-E_{nk_L}),
\end{equation}
where $T_{nk_L}=\frac{h}{L_z} v_{nk_L}=\frac{h}{L_z} (\frac{\hbar k_L}{m})$ and $k_L$ denotes the subset of right-moving wavevectors, $k_L \equiv \{ k|k>0  \}$. Applying the substitution $\sum_{k_L} \mapsto \frac{L_z}{2\pi} \int_0^{\frac{\pi}{L_z}} dk$, the transmission coefficient becomes:
\begin{equation}
    \tau_{n}(E)=\frac{L_z}{2\pi} \int  _{0}^{\frac{\pi}{L_z}} \frac{h}{L_z} \frac{\hbar k}{m}  \delta(E-E_{nk})  dk.
\end{equation}

We define $f(k) \equiv \delta[E-E_{nk}]$ such that $\delta[E-E_{nk}]=\delta[f(k)]$. For each energy $E$ with $E>E_{nC}$, the equation $f(k)=0$ has two solutions: 
$k_{n}^{-}(E)=-\sqrt{2m(E-E_{nC})/\hbar^2}$ and $k_{n}^{+}(E)=\sqrt{2m(E-E_{nC})/\hbar^2}$. Using the property $\delta[f(k)]=\sum_{k_i}\frac{\delta(k-k_i)}{|f'(k_i)|}$, and noting that $|f'[k_{n}^{\pm}(E)]|=\frac{\hbar^2 k_{n}^{\pm}(E)}{m} $, we evaluate the integral as follows:
\begin{align*}
\tau_n(E) &= \frac{L_z}{2\pi} \int_{0}^{\frac{\pi}{L_z}}
\left(\frac{h}{L_z} \frac{\hbar k}{m} \right)
\\
&\cdot \frac{m}{\hbar^2 k_{n}^{+}(E)} \{\delta [k-k_{n}^{-}(E)] + \delta [k-k_{n}^{+}(E)]\}dk
\\
&=\frac{L_z}{2\pi}\left[\frac{h}{L_z} \frac{\hbar k_{n}^{+}(E)}{m} \right]\frac{m}{\hbar^2 k_{n}^{+}(E)}.
\label{tauE-temp}    
\end{align*} 

Similarly, the partial density of states (PDOS) for right-moving electrons in the $n$-th band is:
\begin{equation*}
D_{nk_L}(E) \equiv \sum_{k_L} \delta[E-E_{nk_L}]=\frac{mL_z}{2\pi \hbar^2 k_{n}^{+}(E)}, 
\end{equation*}
where $k_{n}^{+}(E)=\sqrt{2m(E-E_{nC})/\hbar^2}$. Note that the density of states is inversely proportional to velocity \cite{Grosso_ch1}.

Finally, the transmission coefficient simplifies to:
\begin{equation}  
\tau_n(E)= T_{nk_{n}^{+}}(E) \cdot D_{nk_L}(E)=1, 
\label{eqn:tauE-delta}
\end{equation}
where $T_{nk_{n}^{+}(E)} = \frac{h}{L_z}\frac{\hbar k_{n}^{+}(E)}{m}$. 

The energy $E$ intersects the $n-$th energy band $E_{nk}$ at $k=k^{+}_n(E)$, which belongs to the subset $k_L \equiv \{k|k>0\}$, representing the electrons incident from the left. The corresponding transmission coefficient $\tau_n(E)$ is determined by $T_{nk_n^{+}(E)}$, representing the transmission probability density $T_{nk}(E)$ for $k=k_n^{+}(E)$, weighted by the PDOS $D_{nk_L}(E)$, as shown in Eq.~(\ref{eqn:tauE-delta}).

% Add the following section to include some informative discussion in the Reply for the version of resubmission
\section{Additional comparison of NEGF-LCAO and EMT-PW}\label{app:accuracy}
\begin{figure} [ht]
\includegraphics[width=0.9\linewidth]{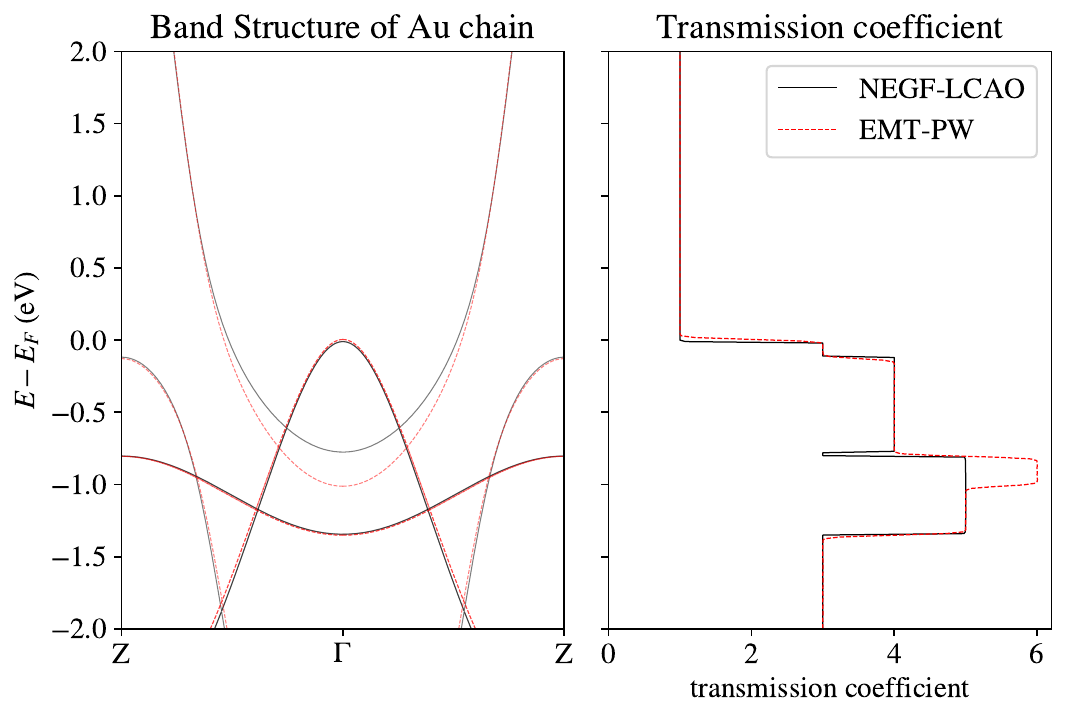}
\caption{\label{fig:Au_wire} 
(color online)
The left panel compares the band structures calculated using NanoDCAL with an LCAO basis (black line) and VASP with a plane-wave (PW) basis (red line). The right panel shows the transmission coefficients computed using NEGF-LCAO based on NanoDCAL (black line) and EMT-PW based on wavefunctions obtained from VASP (red line).
}
\end{figure}

Our EMT-PW method is based on wavefunctions computed using VASP with a plane-wave basis set. Its accuracy is inherently limited to the small-bias regime and depends on the precision of the underlying VASP calculations. This accuracy can typically be enhanced by increasing the number of plane waves and k-points. In contrast, NanoDCAL employs NEGF-DFT with a linear combination of atomic orbitals (LCAO), which constitutes a finite basis set. Unlike the plane-wave basis, which is a complete set, atomic orbitals must be carefully chosen for each specific system to ensure both accuracy and computational efficiency. Although NanoDCAL is fully ab initio, often highly accurate, and computationally efficient, this efficiency is achieved at the expense of generality and, in some cases, accuracy.

Unlike the Al atomic chain, where aluminum is a simple s-p metal, the Au chain consists of gold atoms with a more complex electronic configuration, resulting in significant s-d orbital mixing. We compute and compare the electronic structures and transmission coefficients of the Au atomic chain at zero bias using EMT-PW (within the VASP framework) with those from NEGF-DFT (within the NanoDCAL framework). The comparison is presented in Fig.~\ref{fig:Au_wire}.

\begin{figure} [h]
\includegraphics[width=0.9\linewidth]{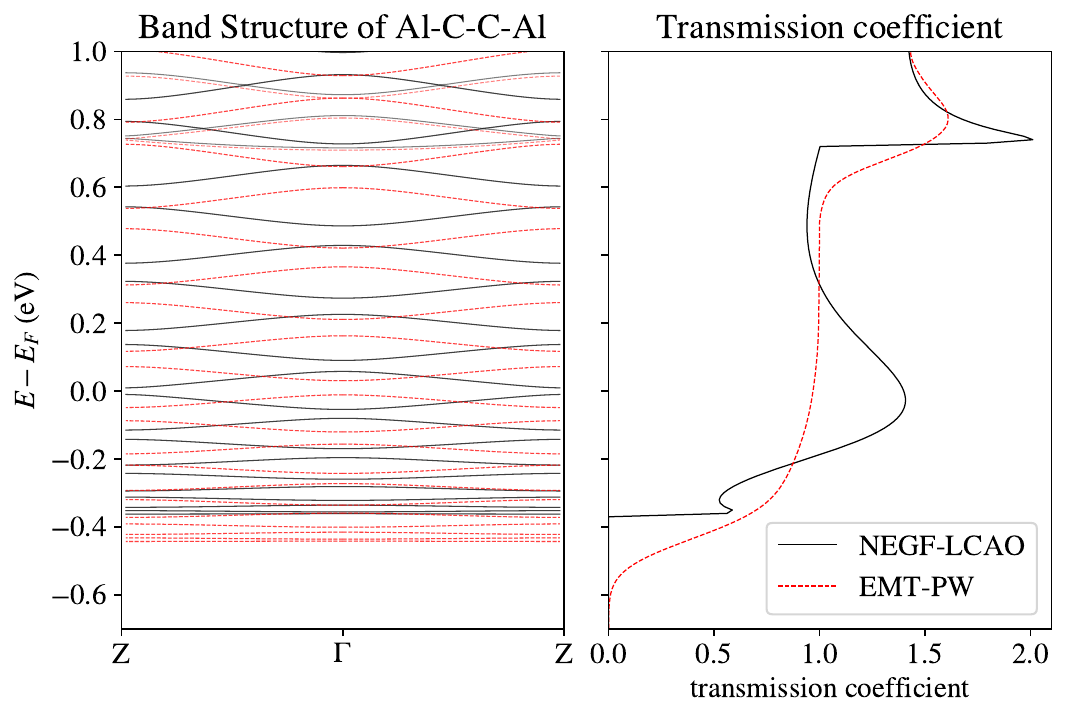}
\caption{\label{fig:Al-CC-Al} 
(color online)
The left panel compares the band structures calculated using NanoDCAL with an LCAO basis (black line) and VASP with a plane-wave (PW) basis (red line). The right panel shows the transmission coefficients computed using NEGF-LCAO based on NanoDCAL (black line) and EMT-PW based on wavefunctions obtained from VASP (red line).
}
\end{figure}

Figure~\ref{fig:Au_wire} (left panel) shows that the band structures obtained using VASP with a plane-wave (PW) basis and NanoDCAL with a linear combination of atomic orbitals (LCAO) differ more significantly than in the case of the Al wire [Fig.~\ref{fig:result}(a)], especially in the conduction band that intersects the Fermi energy ($E_F = 0$). Notably, the conduction band edge obtained from the VASP calculation is lower than that from NanoDCAL. This difference in the band structures results in a corresponding disagreement in the transmission function, as illustrated in the right panel. 

A similar discrepancy in the transmission function arising from differences in the band structure also appears in the Al–C–C–Al structure. As shown in Fig.~\ref{fig:Al-CC-Al}, at $E_F=0$, the band structure obtained from NanoDCAL (LCAO) exhibits a steeper slope compared to that from VASP (PW). Upon closer inspection, one can observe that the small band gaps in the NanoDCAL result are narrower in this region. This difference explains why the transmission function calculated with VASP displays a plateau at $E_F=0$, while that from NanoDCAL does not.

The essence of the EMT (Effective Medium Theory) is that every material inherently carries current—--this current is a fundamental property of its electronic structure. In the absence of an external electric field, however, there is no net current because the left- and right-going currents cancel each other out. The transmission coefficient corresponding to this canceled current can be extracted directly from the electronic structure calculations.

\begin{acknowledgments}
The authors thank MOE ATU, NCHC,  National Center for Theoretical Sciences(South), and NSTC (Taiwan) for support under Grant NSTC 111-2112-M-A49-032-. also supported by NSTC T-Star Center Project: Future Semiconductor Technology Research Center under NSTC 114-2634-F-A49-001-. This work was financially supported under Grant No. NSTC-113-2112-M-A49-037-, and supported by NSTC T-Star Center Project: Future Semiconductor Technology Research Center under NSTC 114-2634-F-A49-001-, and also supported in part by the Ministry of Science and Technology, Taiwan. We thank to National Center for High-performance Computing (NCHC) for providing computational and storage resources.
\end{acknowledgments}

\bibliography{refs}
\end{document}